\documentstyle[12pt]{report}
\begin{document}

\newcommand{\E}{{\cal E}}
\newcommand{\alr}{\alpha r}
\newcommand{\asr}{\alpha^{2}r^{2}}
\newcommand{\dspst}{\displaystyle}

\title{\bf Rotating Charged Black Strings in General Relativity}
\author{Jos\'e P.S. Lemos \thanks{e-mail:lemos@on.br} 
\\
Departamento de Astrof\'{\i}sica\\
Observat\'orio Nacional -- CNPq\\
Rua General Jos\'e Cristino 77,\\
20921, Rio de Janeiro, RJ, Brazil,\\
\& Departamento de F\'{\i}sica\\
Instituto Superior T\'ecnico\\
Av. Rovisco Pais 1,\\
1096, Lisboa, Portugal,\\
\\
Vilson T. Zanchin \thanks{e-mail: zanchin@het.brown.edu}
\\ 
Departamento de F\'{\i}sica-CCNE\\ 
Universidade Federal de Santa Maria \\
97119-900 Santa Maria, RS, Brazil, \\ 
\& Department of Physics \\
Brown University,\\ 
Providence, RI 02912.}
\date{}
\maketitle
\begin{abstract}
\baselineskip 0.6cm
\noindent 
Einstein-Maxwell equations with a cosmological constant are analyzed in 
a four dimensional stationary spacetime admitting in addition a 
two dimensional group $G_2$ of spatial isometries.  We find charged 
rotating black string solutions. For open black strings the mass ($M$), 
angular momentum ($J$) and charge ($Q$) line densities can be  defined 
using the Hamiltonian formalism of Brown and York. It is shown through 
dimensional reduction that $M$, $J$ and $Q$ are respectively the mass, 
angular momentum and charge of a related three dimensional black hole. 
For closed black strings one can define the total mass, charge and 
angular momentum of the solution. These closed black string solutions 
have flat torus topology.  The black string solutions are classified 
according to the mass, charge and angular momentum parameters. The 
causal structure is studied and some Penrose diagrams are shown. There 
are similarities between the charged rotating black string  and the 
Kerr-Newman spacetime. The solution has Cauchy and event horizons, 
ergosphere, timelike singularities, closed timelike curves, and 
extremal cases.  Both the similarities and diferences of these black 
strings and Kerr-Newman black holes are explored.  We comment on the 
implications these solutions might have on the hoop conjecture. 
\\

\end{abstract}

\addtocounter{chapter}{1}
\vspace*{1cm}

\baselineskip 0.9cm
{\bf I. Introduction}

The theory of gravitational collapse and the theory of black holes are 
two distinct but linked subjects. From the work of Oppenheimer and 
Snyder \cite{oppenheimer} and Penrose's theorem \cite{penrose1} we know 
that if General Relativity is correct, then realistic, slightly 
non-spherical, complete collapse leads to the formation of a black hole 
and a singularity. There are also studies hinting  that the introduction 
of a cosmological constant does not alter this picture \cite{hiscock}.

On the other hand, highly non-spherical collapse is not so well
understood.  The collapse of prolate spheroids, i.e. spindles, is not
only astrophysically interesting but also important to a better
understanding of both the cosmic censorship \cite{penrose2} and hoop
\cite{thorne1} conjectures.  Prolate collapse has been studied in some
detail \cite{thorne1,thorne2,echeverria,teukolsky} and it 
was shown that fully relativistic effects,
totally different from the spherical case, come into play. 

Collapse of cylindrical systems and other idealized models was used 
by Thorne to mimic prolate collapse \cite{thorne1}. This study led to 
the formulation of the hoop conjecture 
which states that horizons form when and only when a mass gets compacted 
into a region whose circunference in every direction is less than its 
Schwarzschild circunference, $4\pi GM$ (the velocity of light is 
equal to one in this paper). Thus, 
cylindrical collapsing matter will not form a black hole. However, the 
hoop conjecture was given for spacetimes with zero cosmological 
constant. In the presence of a negative cosmological constant 
one can expect the ocurrence of major changes. Indeed, we show in the 
present work that there are black hole solutions with cylindrical 
symmetry if a negative cosmological constant is present (a fact that does 
not happen for zero cosmological constant). These cylindrical 
black holes are also  called 
black strings.  We study the charged rotating black string and show 
that apart from spacetime being asymptotically anti-de Sitter 
in the radial direction (and 
not asymptotically flat) the black string solution has many similarities 
with the Kerr-Newman black hole. The existence of black strings 
suggests that they can form from the collapse 
of matter with cylindrical symmetry. This hint can indeed be 
confirmed as we next show when we take into account the three dimensional 
(3D) black hole of Ba\~nados, Teitelboim and Zanelli (BTZ) \cite{btz}. 

In  pure (zero cosmological constant) 
3D General Relativity the theory of gravitational collapse has some
features, which like cylindrical collapse in four dimensions (4D), do
not resemble at all 4D spherical collapse. For instance, the radius of
a static fluid 1-sphere does not depend on the pressure, it depends
solely on the energy density of the fluid. In fact, in pure 3D Einstein's
gravity there is no gravitational attraction, and so there is no
gravitational collapse \cite{giddings,zanelli}, essentially
because the theory has no local degrees of freedom and no Newtonian
limit.  Without cosmological constant there is no analogue in 3D of the
Kerr and Schwarzschild black holes spacetimes. However, with a negative
cosmological constant the situation changes. It has been shown that a
black hole exists \cite{btz,bhtz}. This is a surprising result,  
since the black hole spacetime has constant curvature.
However, the solution is only locally anti-de Sitter, but globally has
the topology of a black hole. The rotating case resembles the Kerr
metric and the non-rotating case the Schwarzschild solution, although
there is no polynomial singularity, only a causal singularity.
It has also been shown \cite{mannross} that gravitational collapse in 3D 
with $\Lambda<0$ leads to a black hole. This result connects the 
theory of gravitational collapse with the theory of black holes in 3D.

It is interesting to discuss further the common points between 4D
cylindrical General Relativity and other 3D theories 
\cite{lemos1,lemos2}.  It is well known that straight infinite cosmic string
dynamics and the dynamics of point particles in 3D General Relativity
coincide. Thus, results in 3D point particle General Relativity can be
directly translated into results in straight cosmic string theory (a special
case of cylindrical symmetry in General Relativity \cite{vilenkin}).
The 3D point particle is a solution of the the 3D theory given by
Einstein's action, 
\begin{equation}
S = \frac{1}{16\pi G}\int d^3x \sqrt{-g} (R-2\Lambda)
+S_{\rm f}\, ,
                            \label{eq:11}
\end{equation}
where $S_{\rm f}$ is the action for other fields that might be present,  
$R$ is the curvature scalar, $g$ is the determinant of the metric, 
$\Lambda$ is the cosmological constant and in the 3D case $G$ is usual taken 
as $G=\frac18$, although for the moment we leave it free. 
The usual point particle solution 
appears for zero $\Lambda$ with appropriate matter stress-energy density. 
The action  (\ref{eq:11}) with $\Lambda<0$  yields the 3D BTZ black hole. 
Now the question that we can 
ask is how can we relate a 3D theory, in particular 3D Einstein's gravity, 
with 4D cylindrical General Relativity. By the well known dimensional 
reduction procedure one is able to connect both theories. 
A 4D metric, $g_{\mu\nu}$ ($\mu\nu=0,1,2,3$), with one Killing 
vector can be written (in a particular instance) as, 
\begin{equation}
ds^2 = g_{\mu\nu}dx^\mu dx^\nu = 
g_{mn} dx^m dx^n + {\rm{e}}^{-4\phi}dz^2\, ,
                            \label{eq:12}
\end{equation}
where $g_{mn}$ 
and $\phi$ are metric functions, $m,n=0,1,2$ and $z$ is the 
Killing coordinate. Equation (\ref{eq:12}) is invariant under 
$z\rightarrow-z$. A cylindrical symmetric metric can then be taken from 
(\ref{eq:12}) by imposing that the azimuthal coordinate, $\varphi$,  
also yields a Killing direction. Einstein-Hilbert 
action in 4D
\begin{equation}
S = \frac{1}{16\pi G}\int d^4x \sqrt{-g} (R-2\Lambda)\, ,
                            \label{eq:13}
\end{equation}
plus metric (\ref{eq:12}) give through dimensional reduction 
the following 3D theory:
\begin{equation}
S = \frac{1}{16\pi G}\int d^3x \sqrt{-g} {\rm{e}}^{-2\phi}(R-2\Lambda).
                            \label{eq:14}
\end{equation}   
In equations (\ref{eq:13}) and (\ref{eq:14}) $R$ refers to the 4D and 3D
Ricci scalars, respectively.  In general $\phi$ is not zero (or a
constant). However, the special case $\phi=0$ yields 3D General
Relativity, given in equation (\ref{eq:11}). 
Thus, cylindrical symmetry in which the
Killing direction, $dz$, has got a constant as associated metric function, 
is related directly with 3D General Relativity. If the
energy-momentum tensors in the 4D and 3D theories are also
appropriately related then 4D cosmic strings and 3D point particles
have the same dynamics. In the same way, by a careful choice of an
appropriate energy-momentum tensor in 4D, the 3D BTZ black hole can be
associated with a black string in 4D \cite{kaloper,lemos3}.  
Thus, one can now
translate the results in 3D gravitational collapse with negative
cosmological constant \cite{zanelli,mannross} to 4D cylindrical
symmetry and conclude that cylindrical collapse in a negative
cosmological constant background produces black
strings.

We consider here in this paper, the case in which the dilaton 
$\phi$ is not a
constant. We show that the theory also has black holes 
similar to the Kerr-Newman black holes, with a
polynomial timelike singularity hidden behind the event and Cauchy
horizons. When the charge is zero, the rotating solution does not
resemble so much the Kerr solution, the singularity is spacelike hidden
behind a single event horizon. In addition, in the non-rotating
uncharged case, apart from the topology and asymptotics, the solution
is identical to the Schwarzschild solution.  The 4D  black
string metric has a corresponding 3D black hole solution. It is
likely, in view of the arguments given above, that cylindrical collapse
with appropriate matter in a $\Lambda<0$ background will produce the
black strings discussed in this paper. We note that these black
holes have a cylindrical or toroidal event horizon, in contrast with Hawking's
theorem \cite{hawking1} which states that the topology of the event
horizon is spherical. Again, the presence of a negative cosmological
constant alters the situation \cite{tedjac}.

Cylindrical symmetry, as emphasized by Thorne \cite{thorne1}, is an 
idealized situation. It is possible that the Universe we live in 
contains an infinite cosmic string. It is also possible, however  
less likely, that the Universe is crossed by an infinite  
black string. Yet, one can always argue that close enough to a loop 
string, spacetime resembles the spacetime of an infinite cosmic string. 
In the same way, one could argue that close enough to a toroidal  finite 
black hole, spacetime resembles the spacetime of the infinite black 
string. 

In section II we give the equations and the solutions. In section 
III we discuss and find the mass and angular momentum of the 
solutions. In section IV we study the causal structure of 
the charged rotating black string. In section V we study the causal 
structure of the uncharged rotating black string. 
In section VI we discuss other 
solutions. In section VII we study the geodesics. In section VIII 
we relate the 4D black strings with 3D black holes and in section IX 
we present the conclusions. 

\vspace*{1.cm}
\addtocounter{chapter}{1}
\setcounter{equation}{0}
\setcounter{figure}{0}
\vspace{10pt}

{\bf II. Equations and Solutions}

We consider Einstein-Hilbert action in four dimensions with a cosmological
term in the presence of an electromagnetic field. The total action is 
\begin{equation}
S+S_{\rm em} = \frac{1}{16\pi G}\int{d^{4}x\sqrt{-g}(R-2\Lambda)}
 -\frac{1}{16\pi}\int{d^{4}x\sqrt{-g}F^{\mu\nu}F_{\mu\nu}}\, , 
                        \label{eq:21}
\end{equation}
where $S$ was defined in (\ref{eq:13}), and $R$ and $g$ have 
also been defined in the previous section. 
The Maxwell tensor is 
\begin{equation}
F_{\mu\nu} = \partial_\mu A_\nu - \partial_\nu A_\mu, 
                          \label{eq:22}
\end{equation}
$A_\mu$ being the vector potential. 
We study solutions of the Einstein-Maxwell equations with cylindrical 
symmetry. By this we mean spacetimes admitting a commutative two 
dimensional Lie group $G_2$ of isometries. The topology of the two 
dimensional space generated by $G_2$ can be (i) $R\times S^1$, 
the standard cylindrically symmetric model, with orbits 
diffeomorphic either to cylinders or to $R$ (i.e, $G_2=R\times U(1)$),  
(ii) $S^1 \times S^1$ the flat torus $T^2$ model 
($G_2=U(1) \times U(1)$),  and (iii) $R^2$ 
\cite{chrusciel}. We will focus upon (i) and (ii). 
We then choose a
cylindrical coordinate system $(x^0,x^1,x^2,x^3) = 
({t}, r,{\varphi}, z)$ with
$-\infty<{t}<+\infty$, $0\leq r< +\infty$, $-\infty <z<+\infty$,
$0\leq{\varphi}< 2\pi$. 
In the toroidal model  (ii) the range of the coordinate $z$ is 
$0\leq \alpha z< 2\pi$.
The electromagnetic four potential is given
by $ A_{\mu} =-h(r)\delta_{\mu}^{0}$, where $h(r)$ is an arbitrary
function of the radial coordinate $r$. Solving the Einstein-Maxwell 
equations yielded by  (\ref{eq:21}) for a static 
cylindrically symmetric spacetime we find, 
\begin{equation}
ds^{2}=-\left(\asr-\frac{b}{\alr}
 +\frac{c^2}{\asr}
\right)d{{t}}^{2} -\frac{dr^{2}}{\asr-\frac{b}{\alr}
+\frac{c^2}{\asr}} +r^{2}d{{\varphi}}^{2}
 +{\alpha}^{2}r^{2}dz^{2}\
                          \label{eq:23}
\end{equation}
\begin{equation}
h(r) = \frac{2\lambda}{\alr} + {\rm const.}
                          \label{eq:24}
\end{equation}
where $\alpha^2 \equiv -\frac{1}{3}\Lambda$, $c^2\equiv 4G\lambda^{2}$ and
$b$ and $\lambda$ are integration constants. It is easy to show, for instance
using Gauss's law, that $\lambda$ is the linear charge density of the $z$-line,
and $b=4GM$ with $M$ being the mass per unit length of the $z$-line as we 
will show in the next section. Depending on the raltive values of 
$b$ and $c$, metric (\ref{eq:23}) can represent a static black string. 
In the case there is a black string, 
an analysis of the Einstein-Rosen bridge (see e.g. \cite{MTW}) 
of the  solution (\ref{eq:23}) 
shows that spacetime is not simply connected which here 
implies that the first Betti number of the manifold is one, i.e.,  
closed curves encircling the horizon cannot be shrunk to a point. 

There is also a stationary solution that follows from 
equations (\ref{eq:21})  given by 
\begin{eqnarray}
& ds^{2}= -\left\lbrack (\gamma^2-\frac{\omega^2}{\alpha^2})\asr 
-\frac{\gamma^2b}{\alr} + \frac{\gamma^2c^2}{\asr}\right\rbrack dt^2
-\frac{\gamma\omega}{\alpha^3r}(b-\frac{c^2}{\alr})2d\varphi dt+&
\nonumber\\
&+ \left\lbrack (\gamma^2-\frac{\omega^2}{\alpha^2})r^2 + 
\frac{\omega^2b}{\alpha^5r}-\frac{\omega^2c^2}{\alpha^6r^2}
\right\rbrack d\varphi^2 + \frac{dr^2}
{\asr-\frac{b}{\alr}+\frac{c^2}{\asr}} +\asr dz^2\, ; &  \label{eq:25}
\end{eqnarray}
\begin{equation}
 A_{\mu} =- \gamma h(r) \delta_{\mu}^{0} +\frac{\omega}{\alpha^2}h(r)
\delta_{\mu}^{2}\, ,
                           \label{eq:26}
\end{equation}
where $\omega$ and $\gamma$ are constants, 
$h(r) = \frac{2\lambda}{\alr}$, and  
the coordinates have the same range as in the static case. 
Solution (\ref{eq:25}) can represent a stationary black string. If one 
compactifies the $z$ coordinate ($0\leq \alpha z< 2\pi$) one has a closed  
black string. In this case, one can also put the coordinate $z$ to 
rotate. However, this simply represents a bad choice of coordinates. 
One can always find principal directions in which spacetime 
rotates only along one of these ($\varphi$, say) as in  (\ref{eq:25}).  
This also follows from the fact that the first Betti number is one.

For an observer at radial infinity, the standard 
cylindrical spacetime model (with $R\times S^1$ topology) given by
the metric  (\ref{eq:25}) extends uniformly over the infinite $z-{\rm
line}$.  Thus one expects that, as $r\rightarrow\infty$, the total
energy as well as the total charge is infinite.  The quantities that
can be interpreted physically are the mass and charge densities, i.e.,
mass and charge per unit length of the string.  In fact we have already
found above the finite and well defined line charge density (of the
$z$-line) as an integration constant in Einstein-Maxwell equations. 
For the close black string (the flat torus model 
with $S^1\times S^1)$ topology) the total energy and total charge 
are well defined quantities. 
In order to properly define such quantities we use the Hamiltonian
formalism and the perscription of Brown and York \cite{by1}.

Let us recall that spacetime of (\ref{eq:25}) is also asymptotically
anti-de Sitter whose metric we write here for later reference.
\begin{equation}
 ds^2=-\left(\gamma^{2}-\frac{\omega^2}{\alpha^2}\right)\asr dt^{2}+
\frac{dr^2}{\asr} + \left(\gamma^{2}-\frac{\omega^2}{\alpha^2}
\right) r^{2}d\varphi^{2} + \asr dz^{2}\, .
                                 \label{eq:27}
\end{equation}
To have the usual form of the anti-de Sitter metric we choose 
$\gamma^{2}-\frac{\omega^2}{\alpha^2}=1$. 
This is also the background reference space, since metric (\ref{eq:25})
reduces to (\ref{eq:27}) if the black hole is not present.

\vspace*{1cm}
\addtocounter{chapter}{1}
\setcounter{equation}{0}
\setcounter{figure}{0}
\vspace{10pt}

{\bf III. Mass, angular momentum and charge of the black string}

To use Brown and York formalism let
us write the metric (\ref{eq:25}) in the suitable canonical form
\begin{equation}
ds^2 = -{N^{0}}^2 dt^{2} + R^2(N^{\varphi}dt +d\varphi)^{2} +
\frac{dR^{2}}{f^2} + e^{-4\phi}dz^{2} , 
                                    \label{eq:31}
\end{equation}
where
\begin{eqnarray}
&{N^{0}}^2 =  \left(\gamma^2 - \frac{\omega^{2}}{\alpha^{2}}\right)^2
(\asr-\frac{b}{\alr}+\frac{c^2}{\asr})
\frac{r^{2}}{R^{2}}\, , \hspace*{0.1cm}
N^{\varphi} = - \frac{\gamma\omega}{\alpha^{2}R^{2}}\left(\frac{b}
{\alr} - \frac{c^2}{\asr}\right)\, , &\nonumber \\
& R^2=  \gamma^{2}r^{2} - \frac{\omega^{2}}{\alpha^{4}}
(\asr-\frac{b}{\alr}+\frac{c^2}{\asr})\, ,\hspace*{0.1cm}
f^2 = (\asr-\frac{b}{\alr}+\frac{c^2}{\asr})
\left(\frac{dR}{dr}\right)^2  , &\nonumber \\
&e^{-4\phi} = \asr. &
                                   \label{eq:32}
\end{eqnarray}
In metric (\ref{eq:31}) $N^{0}$ and $N^{\varphi}$ are respectively the
lapse and shift functions. We then choose a region $\cal{M}$ of 
spacetime bounded
by $R={\rm {\rm constant}}$ and two  space-like 
hypersurfaces $t=t_{1}$ and $t=t_{2}$. The
surface $t={\rm constant}$, $R={\rm constant}$ is the two-boundary 
$B$ of the three-space
$\Sigma$. The three-boundary of $\cal{M}$ ($^{3}B$ according to 
Brown and York
symbology) is in the present case the product of $B$ with timelike lines
($R={\rm constant }\, ,\varphi={\rm constant}\, ,z={\rm constant}$).
Then, $B$ can be thought also as the intersection of $\Sigma$ with $^{3}B$.
Notice that, since we are now treating the standard cylindrical symmetric 
model with $R\times S^1$ topology, 
$B$ is an infinite cylindrical surface with radius $R$ and 
infinite total area. Then, in order to avoid
infinite quantities during calculations we select just a finite region of
such a surface, which we call $B_z$. We choose $B_z$ to be  between $z=z_{1}$ 
and $z=z_{2}$.  In the present case the metric $\sigma_{ab}$ can be obtained 
from (\ref{eq:31}) by puting $dt=0$ and $dR=0$ (thus, $a,b=2,3$), 
while the three-space 
metric $h_{ij}$ is obtained by puting $dt=0$ (thus, $i,j=1,2,3$). 
The two-metric on $B$, 
$\sigma_{ab}$, can also be viewed as a spacetime tensor 
$\sigma_{\mu\nu}$ as well as a tensor on the three-space, $\sigma_{ij}$. 

Now we see that the metric given in equation (\ref{eq:31}) admits the
two Killing vectors needed  in order to define mass and angular
momentum: a timelike Killing vector 
$\xi_{t}^{\mu}=(\frac{\partial}{\partial t})^{\mu}$ and a spacelike
(axial) Killing vector 
$\xi_{\varphi}^{\mu}=(\frac{\partial}{\partial \varphi})^{\mu}$.  
Brown and York arrived at the
following definition of a global charge $Q_{\xi}$ related
to any given Killing vector $\xi^{\mu}$:\\ 
\begin{equation}
Q_{\xi} = \int_{B} d^{2}x
\sqrt{\sigma}\left(\epsilon u^{\mu} +j^{\mu} \right)\xi_{\mu} \, ,   
                               \label{eq:33}
\end{equation}         
where $u^{\mu}$ is the timelike future pointing normal to
$\Sigma$, $\sigma$ is the determinant of $\sigma_{ab}$, $\epsilon$ and
$j^{\mu}=(0,j^{i})$ are respectively the energy and momentum 
surface density on $B$, and are given by:
\begin{eqnarray}
\epsilon =\frac{k}{8\pi G}\, , &  & j^{i} = 
\frac{\sigma^{i}_{j}n_{k}\Pi^{jk}}{16\pi G\sqrt{h}} \, ,
\end{eqnarray}
where $n^k$ is the unit normal to the two-boundary $B$ on the 
three-space $\Sigma$, $k$ 
is the trace of the extrinsic
curvature of $B$, $\Pi^{ij}$ is the conjugate momentum in $\Sigma$ 
and $h$ is the determinant of $h_{ij}$. 

In (\ref{eq:33}) $\epsilon$ and $j^{i}$ are defined in such a way that they
vanish for the background (reference) spacetime. The meaning of such
reference spacetime can be seen as follows. Suppose one is
interested in the energy (mass) of a segment of spacetime and tries
to calculate it using Hamiltonian formalism with the gravitational action
$S_1$ obtained
from (\ref{eq:31}).  The result contains also the
contribution from the background spacetime which in our case is the anti-de
Sitter spacetime whose metric is given by (\ref{eq:27}), or equivalently, by
(\ref{eq:31}) with $b=0$ and $c=0$.  Such a ``background" spacetime energy 
does not vanish
even in the absence of the black string, and it diverges at 
radial infinity. Then, in order to properly define the energy of the black 
string one must get rid of  
the contribution of the anti-de Sitter background
spacetime to the total energy of the black string spacetime. 
Let us then define $S_0$ as the action related to the reference spacetime,
while $S_1$ is the full gravitational action of the black string spacetime 
obtained from metric (\ref{eq:31}). Both, $S_1$ and $S_0$ are functionals of
the lapse and shift functions suitable defined on the boundary $^3B$
\cite{by1,by2}.  Thus, from $S_1$ we get the total surface energy density
at a given point $P$ on the boundary $B$ as 
$\epsilon_{1} = k_{1}/8\pi G$ where
$k_{1}$ is the trace of the extrinsic curvature of $B$, obtained from
(\ref{eq:31}), when the
black string is present. In a similar way, from $S_0$ we get
$\epsilon_{0} = k_{0}/8\pi G$,
the background surface energy density at the same 
point $P$ on $B$. $k_{0}$ is
the trace of the extrinsic curvature obtained 
from (\ref{eq:31}) with $b=0$ and $c=0$. 

Now, we see that the surface energy density $\epsilon=\epsilon_{1}- 
\epsilon_{0}$ defined at each point on the boundary $B$ is in fact the 
contribution to energy density due only to the presence of the black 
string, since we are subtracting the contribution of the background 
reference space from the total energy density of the black string  
spacetime. Notice also that this ``true" energy density (of the black 
string) $\epsilon$ may be obtained from the ``true" gravitational 
action $S = S_1-S_0$ defined on the boundary $^3B$. 

It is all the same regarding to, for instance, the angular momentum of 
the black string spacetime. In order to properly define the angular 
momentum, the background surface momentum density $j_{0}^{i}$ must be 
subtracted from the ``total" surface density  of the black string 
spacetime $j_{1}^{i}$. Thus, in (\ref{eq:33}) we have to take, 
$\epsilon = (k_{1}-k_{0})/8\pi G$ and $j^{i}= j_{1}^{i}-j_{0}^{i}$. 
 
Then, equation (\ref{eq:33}) yields the following expressions for 
the total energy and angular momentum of a
segment $\Delta z$ of the two-boundary, $B_z$, at radial infinity
($R\rightarrow\infty$), of the black string spacetime:  
$M_{t}= \dspst{\frac{\Delta z}{8G}b\left(2\gamma^{2} +
\frac{\omega^2}{\alpha^2}\right)}\,$, $J_{\varphi}=\dspst{\frac{3\Delta
z} {8G}b\frac{\gamma\omega}{\alpha^2}}\,$. From these we define 
the mass and angular momentum line densities of the spacetime as,
\begin{equation}
M\equiv\frac{M_{t}}{\Delta z}=\frac{b}{8G}\left(2\gamma^{2}
+\frac{\omega^2} {\alpha^2}\right)\; ,
                                \label{eq:34}
\end{equation}
\begin{equation}
J\equiv \frac{J_{\varphi}}{\Delta z} = \frac{3b}{8G}
 \frac{\gamma\omega}{\alpha^2}\,  .   
                                \label{eq:35}
\end{equation}
Then one can solve a quadratic equation for $\gamma^2$ and
$\frac{\omega^2}{\alpha^2}$. It gives two distinct solutions
\begin{equation}
\gamma^2 = \frac{2GM}{b}+ \frac{2G}{b}\sqrt{M^2 - \frac{8J^2\alpha^2}{9}}
 \quad;\quad
\frac{\omega^2}{\alpha^2} = \frac{4GM}{b}- \frac{4G}{b}\sqrt{M^2 -
 \frac{8J^2\alpha^2}{9}}
                                      \label{eq:36}
\end{equation}
\begin{equation}
\gamma^2 = \frac{2GM}{b} -\frac{2G}{b}\sqrt{M^2 - \frac{8J^2\alpha^2}{9}}
\quad;\quad
\frac{\omega^2}{\alpha^2} = \frac{4GM}{b} + \frac{4G}{b}\sqrt{M^2 - 
\frac{8J^2\alpha^2}{9}}
                                       \label{eq:37}
\end{equation}
We will concentrate on (\ref{eq:36}). 

One can also work out the total 
energy $M_{\rm c}$ and angular momentum $J_{\rm c}$ 
for the closed black string. 
In terms of $M$ and $J$ they are given by $M_{\rm c}=\frac{2\pi}{\alpha} {M}$ 
and $J_{\rm c}=\frac{2\pi}{\alpha} J$. 

Now we turn our attention to the electric charge of the rotating black hole.
In the Hamiltonian formalism the canonical field variables are the spatial components 
of the vector potential  (described by the 1-form $A_{\mu}dx^{\mu}$) 
and the conjugate momentum ${\E}^{i}$, while the
component $A_{0}$ is the Lagrange multiplier for the Gauss-Law constraint.
Notice that ${\E}^{i}$ can be viewed as a tensor in the three-space $\Sigma$.
The electromagnetic field gives rise to boundary terms in the action so that
the electric charge $Q_{z}$ is defined from them as the conjugate
quantity to $A_{0}$ in the action \cite{by3}:
\begin{equation}
Q_{z}= \frac{1}{4\pi}\int_{B_z} d^{2}x \frac{n_{i}{\E}^{i}\sqrt\sigma}
{\sqrt{h}}\, , \hspace{.5cm}
 {\E}^{i}=\frac{R\, e^{-2\phi}}{N}\left({A_{0}}_{,R} - N^{\varphi}
  {A_{2}}_{,R}\right), 
                                 \label{eq:38}
\end{equation}
where we have suppressed the term related to the reference space because it
vanishes since ${\E}_{0}^{i}$ is identically zero. The quantity
$n_{i}{\E}^{i}$ can also be viewed as a surface charge density on the
two boundary $B$.

Then, using (\ref{eq:38}) and the vector potential given in (\ref{eq:26}) we
 find (recall that $B_z$ is the surface of a cylinder whose
radius is $R\rightarrow\infty$ and height is 
$\Delta z$) $Q_{z}= \gamma\lambda\Delta z$. Again
we see that the total charge is in fact infinite, but the line charge density
of the string, which is the quantity that appears in the metric,
\begin{equation}
Q \equiv \frac{Q_{z}}{\Delta z}= \gamma\lambda \, , 
                          \label{eq:39}
\end{equation}
is finite and well defined.The electric charge of the closed black string 
is $Q_{\rm c}=\frac{2\pi}{\alpha} Q $. \\

Finally, using (\ref{eq:36}) and (\ref{eq:39}),  the metric (\ref{eq:25})
assumes the form
\begin{eqnarray}
&ds^{2}= -\left(\asr-\frac{2G(M+\Omega)}{\alr} +\frac{4GQ^2}{\asr}\right)dt^{2}
-\frac{16GJ}{3\alr}\left(1-\frac{2Q^2}{(M+\Omega)\alr}\right) dt d\varphi &
\nonumber \\
&+\left[r^{2}+{4G(M-\Omega)\over{\alpha^{3}r}}\left(1-{2\over{(M+\Omega)}}
\frac{Q^2} {\alr}\right)\right]d\varphi^{2}&\nonumber \\
& +{dr^{2}\over{\asr -\frac{2G(3\Omega-M)}{\alr}+\frac{3\Omega-M}
{\Omega+M}
 \frac{4GQ^2}{\asr}}}+{\asr} dz^{2}\, ,&
                             \label{eq:310}
\end{eqnarray}
where 
\begin{equation}
\Omega =\sqrt{M^2-\frac{8J^2\alpha^2}{9}}.
                             \label{eq:311}
\end{equation}
The asymptotic group of this metric, as $r\rightarrow\infty$, is 
$R\times$conformal group in two dimensions \cite{brownhenneaux}.

\vspace*{1.cm}
\addtocounter{chapter}{1}
\setcounter{equation}{0}
\setcounter{figure}{0}
\vspace{10pt}

{\bf IV. Causal Structure of the Charged Rotating Black String Spacetime}

In order to study the metric and its causal structure it is 
useful to define the parameter $a$  (with units of angular momentum 
per unit mass), 
\begin{equation}
a^2\alpha^2 \equiv 1-\frac{\Omega}{M}                      
                            \label{eq:4001}
\end{equation}
such that 
\begin{equation}
1+\frac{\Omega}{M}=2(1-\frac{a^2\alpha^2}{2})\quad,\quad 
3\frac{\Omega}{M}-1
=2(1-\frac{3}{2}a^2\alpha^2).
                            \label{eq:4002}
\end{equation}
The relation between $J$ and $a$ is given by
\begin{equation}
J=\frac{3}{2}aM\sqrt{1-\frac{a^2\alpha^2}{2}}.
                          \label{eq:4003}
\end{equation}
The range of $a$ is $0\leq a\alpha\leq1$. From now on we put $G=1$ 
(note that in 
\cite{lemos1,lemos2} we have normalized differently, with $G=\frac18$). 
With these definitions the metric (\ref{eq:310}) assumes the form 
\begin{eqnarray}
&ds^{2} = -\left(\asr -\frac{4M(1-\frac{a^2\alpha^2}{2})}{\alr} + 
\frac{4Q^2}{\asr}\right) dt^2 + &
\nonumber \\
&-\frac{4aM\sqrt{1-\frac{a^2\alpha^2}{2}}}{\alr}\left(1-
\frac{Q^2}{M(1-\frac{a^2\alpha^2}{2})\alr}\right) 2dt d\varphi +&
\nonumber \\
&+ \left(\asr -\frac{4M(1-\frac{3}{2}a^2\alpha^2)}{\alr} + 
\frac{4Q^2}{\asr} 
\frac{(1-\frac{3}{2}a^2\alpha^2)}{(1-\frac{a^2\alpha^2}{2})}
\right)^{-1} dr^2 +&
\nonumber \\
&+ \left[r^2 + \frac{4Ma^2}{\alr}\left(1-
\frac{Q^2}{(1-\frac{a^2\alpha^2}{2})M\alr}\right)\right] d\varphi^2 + 
\asr dz^2.&
                             \label{eq:4004}
\end{eqnarray}
In order to compare metric (\ref{eq:4004}) with the well-known 
Kerr-Newman metric we write explicitly here the Kerr-Newman metric 
on the equatorial plane 
\begin{eqnarray}
& ds^2 = -(1-\frac{2m}{r}+\frac{e^2}{r^2}) dt^2 +&
\nonumber \\
&-\frac{2ma}{r}(1-\frac{e^2}{2mr}) 2dtd\varphi+&
\nonumber \\
&+(1-\frac{2m}{r}+\frac{a^2+e^2}{r^2})^{-1} dr^2 +&
\nonumber \\
&+ \left\lbrack r^2 +a^2(1+\frac{2m}{r}-\frac{e^2}{r^2})\right\rbrack 
d\varphi^2 + r^2d\theta^2,&
                             \label{eq:4005}
\end{eqnarray}
where $(m,a,e)$ are the Kerr-Newman parameters, mass, specific angular
momentum and charge, respectively. We can now see that  the metric for
a rotating cylindrical symmetric spacetime asymptotically anti-de
Sitter, given in (\ref{eq:4004}), has many similarities with the
metric on the equatorial plane for an axisymmetric rotating spacetime
asymptotically flat given by the Kerr-Newman metric in (\ref{eq:4005}). 
Of course, there are differences, and we will explore both the differences 
and similarities along this paper.

Metric (\ref{eq:4004}) has a singularity at $r=0$. 
The Kretschmann scalar $K$ is 
\begin{equation}
K={24\alpha^{4} \left(1+ \frac{b^2}{2
\alpha^{6}r^6}\right)- \frac{48 c^2}{\alpha^{3}r^7}\left(b-\frac{7c^2}
{6\alpha\ r}\right)}
                             \label{eq:4006}
\end{equation}
where $b$ and $c$ can be picked up from  (\ref{eq:25}) and 
(\ref{eq:4004}). Thus $K$ diverges at $r=0$. The solution has 
totally different character depending on whether $r>0$ or $r<0$. The 
important black hole solution exists for $r>0$ which case we analyze first.
\vskip 0.5cm

{\bf IV.1}\quad $r\geq0$ (or $M>0$)

To analyze the causal structure and follow the 
procedure of Boyer and Lindquist \cite{boyer} and Carter \cite{carter1} 
we put metric (\ref{eq:4004}) in the form, 
\begin{equation}
ds^{2} = -\Delta 
\left( \gamma dt - \frac{\omega}{\alpha^2} d\varphi\right)^2 + 
r^2 \left(\gamma d\varphi - \omega dt\right)^2 +
\frac{dr^2}{\Delta} +
\asr dz^2\, ,
                             \label{eq:4101}
\end{equation}
where now,
\begin{equation}
\Delta = \asr - \frac{b}{\alr} + \frac{c^2}{\asr}, 
                             \label{eq:4102}
\end{equation}

\begin{equation}
b=4M\left(1-\frac{3}{2}a^2\alpha^2\right),
                             \label{eq:4103}
\end{equation}
\begin{equation} 
c^2= 4Q^2\left( \frac{1-\frac{3}{2}a^2\alpha^2}
{1-\frac12{a^2\alpha^2}}\right),
                             \label{eq:4104}
\end{equation}
\begin{equation}
\gamma=\sqrt{ \frac{1-\frac12{a^2\alpha^2}}{1-\frac{3}{2}a^2\alpha^2} },
                             \label{eq:4105}
\end{equation}
\begin{equation}
\omega= \frac{a\alpha^2}{\sqrt{1-\frac{3}{2}a^2\alpha^2}}.
                             \label{eq:4106}
\end{equation}
There are horizons whenever
\begin{equation}
\Delta=0
                             \label{eq:4107},
\end{equation}
i.e., at the roots of $\Delta$. 
One knows that the non-extremal situations in the Kerr-Newman metric are
given by $0\leq \frac{a^2}{m^2} \leq 1 - \frac{e^2}{m^2}$.
Here, to have horizons one needs either one of the two conditions: 
\begin{equation} 
0\leq a^2\alpha^2\leq \frac23 - \frac{128}{81}
\frac{Q^6}{M^4(1-\frac12a^2\alpha^2)^3},
                            \label{eq:4108}
\end{equation}
or
\begin{equation}
\frac23<a^2\alpha^2\leq1.
                            \label{eq:4109}
\end{equation}
Thus there are five distinct cases depending on the value of the 
charge and angular momentum:
(i) $0\leq a^2\alpha^2\leq \frac23 - \frac{128}{81}
\frac{Q^6}{M^4(1-\frac12a^2\alpha^2)^3}$, which yields the black hole solution 
with event and Cauchy horizons. 
(ii) $a^2\alpha^2 = \frac23 - \frac{128}{81}
\frac{Q^6}{M^4(1-\frac12a^2\alpha^2)^3}$, 
which corresponds to the extreme case, where the two horizons merge. 
(iii)  $\frac23 - \frac{128}{81}
\frac{Q^6}{M^4(1-\frac12a^2\alpha^2)^3} < a^2\alpha^2<\frac23$, corresponding 
to naked singularities solutions. 
(iv) $a^2\alpha^2=\frac23$, which gives a null singularity. 
(v) $\frac23<a^2\alpha^2 <1$, which gives a black hole 
solution with one horizon. 
The most interesting solutions are given in items (i) and (ii). Solutions 
(iv) and (v) do not have  partners in the Kerr-Newman family. 
In figure  4.1, we show the black hole and naked singularity regions, 
and the extremal black hole line dividing those two regions, as well as 
the other solutions in the upper part of the figure. We now analyze 
each item in turn.

\vskip 0.3cm
(i)  $0\leq a^2\alpha^2\leq \frac23 - \frac{128}{81}
\frac{Q^6}{M^4(1-\frac12a^2\alpha^2)^3}$ 

This is the charged-rotating black string spacetime. 
As we will see this one is indeed very similar to the Kerr-Newman black 
hole. The structure has event and Cauchy horizons, timelike 
singularities, and closed timelike curves. 

Now,  following Boyer and Lindquist we choose a new angular coordinate 
which straightens out the helicoidal null geodesics that pile up 
around the event horizon. A good choice is 
\begin{equation}
\overline{\varphi} = \gamma \varphi - \omega t.
                              \label{eq:4110}
\end{equation}
In this case the metric reads, 
\begin{eqnarray}
&ds^{2} = -\left(\asr - \frac{b}{\alr} + \frac{c^2}{\asr} \right) 
\left( \frac{dt}{\gamma} - \frac{\omega}{\alpha^2\gamma} 
d\overline{\varphi}\right)^2 +  r^2 d\overline{\varphi}^2 +&
\nonumber \\
&+ \left(\asr - \frac{b}{\alr} + \frac{c^2}{\asr}\right)^{-1} dr^2 +
\asr dz^2. &
                             \label{eq:4111}
\end{eqnarray}
The horizons are given for the zeros of the lapse function, i.e.
when $\Delta=0$. 
We find that $\Delta$ has two roots $r_+$ and $r_-$ which are given 
by,
\begin{equation}
r_+ = b^\frac13\frac{\sqrt{s} + \sqrt{2\sqrt{s^2-4q^2}-s}}{2\alpha}
                             \label{eq:4112}
\end{equation}
and 
\begin{equation}
r_- =b^\frac13 \frac{\sqrt{s} - \sqrt{2\sqrt{s^2-4q^2}-s}}{2\alpha}
                             \label{eq:4113}
\end{equation}
where,
\begin{equation}
s = \left( \frac12 + \frac12 \sqrt{1-4\left(\frac{4q^2}{3}\right)^3}\right)
^{\frac13} +
 \left( \frac12 - \frac12 \sqrt{1-4\left(\frac{4q^2}{3}\right)^3}\right)
^{\frac13},
                             \label{eq:4114}
\end{equation}
\begin{equation}
q^2 = \frac{c^2}{b^\frac43}
                             \label{eq:4115}
\end{equation}
and $b$ and $c$ are given in equations  (\ref{eq:4103}) and  
(\ref{eq:4104}).

We now introduce a Kruskal coordinate patch around each of the roots 
of $\Delta$, $r_+$ and $r_-$. 
The first patch constructed around $r_+$ is valid for $r_-<r<\infty$.

In the region  $r_-<r\leq r_+$ the null Kruskal coordinates $U$ and $V$ 
are given by,
\begin{eqnarray}
& U = \left( \frac{\alpha\left(r_+ - r\right)}{b^\frac13}\right)^\frac12
\left( \frac{\alpha\left(r - r_-\right)}{b^\frac13}\right)^{-
\frac{C}{B}\frac{{r_-}^2}{2{r_+}^2} } 
F\left( r\right) \exp\left( -\frac{A}{B}\frac{r_+-r_-}{2{r_+}^2} 
\alpha \frac{t}{\gamma}\right)&
\nonumber \\
& V = \left( \frac{\alpha\left(r_+ - r\right)}{b^\frac13}\right)^\frac12
\left( \frac{\alpha\left(r - r_-\right)}{b^\frac13}\right)^{-
\frac{C}{B}\frac{{r_-}^2}{2{r_+}^2} } 
F\left( r\right) 
\exp\left( \frac{A}{B}\frac{r_+-r_-}{2{r_+}^2} \alpha 
\frac{t}{\gamma}\right).  ,&
                             \label{eq:4116}
\end{eqnarray}
For $r_+\leq r<\infty$ we put, 
\begin{eqnarray}
& U = -\left( \frac{\alpha\left(r - r_+\right)}{b^\frac13}\right)^\frac12
\left( \frac{\alpha\left(r - r_-\right)}{b^\frac13}\right)^{-
\frac{C}{B}\frac{{r_-}^2}{2{r_+}^2} } 
F\left( r\right) \exp\left(-\frac{A}{B}\frac{r_+-r_-}{2{r_+}^2} 
\alpha \frac{t}{\gamma}\right)&
\nonumber \\
& V = \left( \frac{\alpha\left(r - r_+\right)}{b^\frac13}\right)^\frac12
\left( \frac{\alpha\left(r - r_-\right)}{b^\frac13}\right)^{-
\frac{C}{B}\frac{{r_-}^2}{2{r_+}^2} } 
F\left( r\right) \exp\left(\frac{A}{B}\frac{r_+-r_-}{2{r_+}^2} 
\alpha \frac{t}{\gamma}\right)&
                             \label{eq:4117}
\end{eqnarray}
The following definitions have been introduced in order to 
facilitate the notation, 
\begin{equation}
A \equiv \left( {r_+}^2 + {r_-}^2\right)^2 + 2\left( r_+ + r_-\right)^4,
                             \label{eq:4118}
\end{equation}
\begin{equation}
B\equiv \frac{1}{\alpha} \left\lbrack \left( r_+ + r_-\right)^2 + 
2{r_-}^2\right\rbrack, 
                             \label{eq:4119}
\end{equation}
\begin{equation}
C \equiv \frac{1}{\alpha} \left\lbrack \left( r_+ + r_-\right)^2 + 
2{r_+}^2\right\rbrack,
                             \label{eq:4120}
\end{equation}
\begin{equation}
D\equiv \frac{1}{2\alpha} \left( r_+ + r_-\right)^3, 
                             \label{eq:4121}
\end{equation}
\begin{equation}
E \equiv \frac{1}{\alpha}
{
{ \left({r_+}^2 + {r_-}^2\right)^2 + 2\left( r_+ + r_-\right)^2
\left( {r_+}^2 + {r_-}^2 + r_+ r_-\right) } \over
{ \sqrt{ \left( r_+ + r_-\right)^2 + 2 \left( {r_+}^2 + {r_-}^2\right)} }
},
                             \label{eq:4122}
\end{equation}
and finally, 
\begin{eqnarray}
&F\left( r\right) \equiv 
\left( \frac{\alpha^2}{b^\frac23}\left\lbrack r^2 + 
\left( r_+ + r_-\right) r + \left( {r_+}^2 + {r_-}^2 + r_+r_-\right)
\right\rbrack \right)^
{-\frac{D}{B} \frac{r_+ - r_-}{2{r_+}^2} }&
\nonumber \\
&\exp\left( 
\frac{E}{B}\frac{r_+ - r_-}{2{r_+}^2}
\arctan \frac{2r + \left( r_+ + r_-\right)}
{\sqrt{ \left( r_+ + r_-\right)^2 + 2 \left( {r_+}^2 + {r_-}^2\right)}}
\right).&
                             \label{eq:4123}
\end{eqnarray}
In this first coordinate patch,  $r_-<r\leq \infty$, the metric can 
be written as,
\begin{eqnarray}
& ds^2 = -\frac{b^\frac23\left\lbrack\alpha\left( r-r_-\right)/b^\frac13
\right\rbrack^{1+\frac{C}{B}\frac{{r_-}^2}{{r_+}^2}}}{\alpha^2{k_+}^2r^2}
G_+\left( r\right) dUdV + &
\nonumber \\
&\frac{a}{\alpha\sqrt{1-\frac{a^2\alpha^2}{2}}}
\frac
{b^\frac23\left\lbrack\alpha\left(r-r_-\right)/b^\frac13\right\rbrack
^{ 1+\frac{C}{B}\frac{{r_-}^2}{{r_+}^2} }}
{k_+ r^2} 
G_+\left( r\right) \left( VdU-UdV\right)d\overline{\varphi} + &
\nonumber \\
&+\left( r^2 - \Delta\frac{a^2}{1-\frac{a^2\alpha^2}{2}} \right)
d\overline{\varphi}^2 + \asr dz^2, &
                             \label{eq:4124}
\end{eqnarray}
where,
\begin{equation}
G_+(r) \equiv \frac{r^2 + \left( r_+ + r_-\right) r + 
\left( {r_+}^2 + {r_-}^2 + r_+r_-\right)}
{F^2(r)}, 
                             \label{eq:4125}
\end{equation}
and
\begin{equation}
k_+ = \frac{A}{B}\frac{r_+ - r_-}{2{r_+}^2}.
                             \label{eq:4126}
\end{equation}
We see that the metric given in (\ref{eq:4124}) is regular in this
patch, and in particular is regular at $r_+$. It is however singular at
$r_-$.  To have a metric non-singular at $r_-$ one has to define new
Kruskal coordinates for the patch $0<r<r_+$.

For $0<r\leq r_-$ we have, 
\begin{eqnarray}
& U = -\left( \frac{\alpha\left(r_+ - r\right)}{b^\frac13}\right)^
{-\frac{B}{C}\frac{{r_+}^2}{2{r_-}^2} } 
\left( \frac{\alpha\left(r_- - r\right)}{b^\frac13}\right)^\frac12
H\left( r\right) \exp\left( \frac{A}{C}\frac{r_+-r_-}{2{r_-}^2} 
\alpha \frac{t}{\gamma}\right), &
\nonumber \\
& V = \left( \frac{\alpha\left(r_+ - r\right)}{b^\frac13}\right)^
{-\frac{B}{C}\frac{{r_+}^2}{2{r_-}^2} } 
\left( \frac{\alpha\left(r_- - r\right)}{b^\frac13}\right)^\frac12
H\left( r\right) \exp\left( -\frac{A}{C}\frac{r_+-r_-}{2{r_-}^2} 
\alpha \frac{t}{\gamma}\right) ,&
                             \label{eq:4127}
\end{eqnarray}
For $r_-\leq r<r_+$ we have, 
\begin{eqnarray}
& U = \left( \frac{\alpha\left(r_+ - r\right)}{b^\frac13}\right)^
{-\frac{B}{C}\frac{{r_+}^2}{2{r_-}^2} } 
\left( \frac{\alpha\left(r - r_-\right)}{b^\frac13}\right)^\frac12
H\left( r\right) \exp\left( \frac{A}{C}\frac{r_+-r_-}{2{r_-}^2} 
\alpha \frac{t}{\gamma}\right), &
\nonumber \\
& V = \left( \frac{\alpha\left(r_+ - r\right)}{b^\frac13}\right)^
{-\frac{B}{C}\frac{{r_+}^2}{2{r_-}^2} } 
\left( \frac{\alpha\left(r - r_-\right)}{b^\frac13}\right)^\frac12
H\left( r\right) \exp\left( -\frac{A}{C}\frac{r_+-r_-}{2{r_-}^2} 
\alpha \frac{t}{\gamma}\right) ,&
                            \label{eq:4128}
\end{eqnarray}
where, 
\begin{eqnarray}
&H(r) = 
\left( \frac{\alpha^2}{b^\frac23}\left\lbrack r^2 + 
\left( r_+ + r_-\right) r + \left( {r_+}^2 + {r_-}^2 + r_+r_-\right)
\right\rbrack \right)^
{\frac{D}{C} \frac{r_+ - r_-}{2{r_-}^2} }&
\nonumber \\
&\exp\left(- 
\frac{E}{C}\frac{r_+ - r_-}{2{r_-}^2}
\arctan \frac{2r + \left( r_+ + r_-\right)}
{\sqrt{ \left( r_+ + r_-\right)^2 + 2 \left( {r_+}^2 + {r_-}^2\right)}}
\right).&
                             \label{eq:4129}
\end{eqnarray}
The metric for this second patch can be written as
\begin{eqnarray}
& ds^2 = -\frac{b^\frac23\left\lbrack\alpha\left( r_+-r\right)/b^\frac13
\right\rbrack^{1+\frac{B}{C}\frac{{r_+}^2}{{r_-}^2}}}{\alpha^2{k_-}^2r^2}
G_-\left( r\right) dUdV + &
\nonumber \\
&-\frac{a}{\alpha\sqrt{1-\frac{a^2\alpha^2}{2}}}
\frac{b^\frac23\left\lbrack\alpha\left(r_+-r\right)/b^\frac13\right\rbrack
^{1+\frac{B}{C}\frac{{r_+}^2}{{r_-}^2}}
}{k_-r^2} G_-\left( r\right) \left( VdU-UdV\right)d\overline{\varphi} + &
\nonumber \\
&+\left( r^2 - \Delta\frac{a^2}{1-\frac{a^2\alpha^2}{2}} \right)
d\overline{\varphi}^2 + \asr dz^2, &
                             \label{eq:4130}
\end{eqnarray}
where,
\begin{equation}
G_-(r) \equiv \frac{r^2 + \left( r_+ + r_-\right) r + 
\left( {r_+}^2 + {r_-}^2 + r_+r_-\right)}
{H^2(r)}
                             \label{eq:43131}
\end{equation}
and
\begin{equation}
k_- = \frac{A}{B}\frac{r_+ - r_-}{2{r_-}^2}.
                             \label{eq:4132}
\end{equation}
The metric is regular at $r_-$ and is singular at $r=0$.  To construct
the Penrose diagram we have to define the Penrose coordinates, $\psi$, $\xi$
by the usual arctangent functions of $U$ and $V$, 
\begin{equation}
U=\tan\frac12(\psi-\xi)\quad \quad V=\tan\frac12(\psi+\xi)
                             \label{eq:4133}
\end{equation}
From (\ref{eq:4133}), (\ref{eq:4116}) and (\ref{eq:4117}) we have that 
in the first patch 
(i) the line $r=\infty$ is mapped into two symmetrical curved  timelike 
lines (ii) the line $r=r_+$ is mapped into two mutual perpendicular
straight lines at $45^{0}$. From (\ref{eq:4127}) and 
(\ref{eq:4128}) we see that (i) $r=0$ is mapped into a curved timelike 
line and (ii) $r=r_-$ is mapped into two mutual perpendicular
straight null lines at $45^{0}$. One has to join these two 
different patches (see \cite{chandra,bhtz}) 
and then repeat them over 
in the vertical. The result is the Penrose diagram shown in figure 4.2. 
The lines $r=0$ and $r=\infty$ are drawn as vertical lines, 
although in the 
coordinates $\psi$ and $\xi$ they should be curved outwards, bulged. 
It is always possible to change coordinates so that the lines are indeed 
vertical.

To study closed timelike curves (CTCs) we first note that the angular Killing 
vector $\frac{\partial}{\partial\varphi}$ has norm given by
\begin{equation}
\frac{\partial}{\partial\varphi}.\frac{\partial}{\partial\varphi} 
= g_{\varphi\varphi} =
r^2 + \frac{4Ma^2}{\alr} - \frac{4a^2Q^2}{(1-\frac{a^2\alpha^2}{2})
\asr}\, .
                             \label{eq:4134}
\end{equation}
There are CTCs for $g_{\varphi\varphi}<0$. One can show that 
the radius for which $g_{\varphi\varphi}=0$ is 
\begin{equation}
r_{\rm CTC} = (\frac{a^2\alpha^2b}{1-\frac32 a^2\alpha^2})^\frac13
\frac{ \sqrt{
2\sqrt
{\overline{s}^2+4\overline{q}^2}-\overline{s} } - 
\sqrt{\overline{s}}   }{2\alpha}
                             \label{eq:4135}
\end{equation}
where,
\begin{equation}
\overline{s} = \left( \frac12 + \frac12 
\sqrt{1+4\left(\frac{4\overline{q}^2}{3}\right)^3}\right)
^{\frac13} +
 \left( \frac12 - \frac12 
\sqrt{1+4\left(\frac{4\overline{q}^2}{3}\right)^3}\right)
^{\frac13},
                             \label{eq:4136}
\end{equation}
\begin{equation}
{\overline{q}}^2 = \frac{a^2\alpha^2}{{1-\frac32 a^2\alpha^2}}
\frac{c^2}{b^\frac43}
                             \label{eq:4137}
\end{equation}
and $b$ and $c$ are given in equations  (\ref{eq:4103})  (\ref{eq:4104}).
One can also show that $r_->r_{\rm CTC}$ always, as in figure  4.2. 
Following 
Carter \cite{carter1} the region $0<r<r_-$ is a vicious region, i.e., 
there are CTCs through every point of it. 

The singularity is at $r=0$ and timelike. Going to Kerr-Schild like 
coordinates near the singularity one can show that the singularity has a 
ring structure, like the Kerr-Newman black hole.
Indeed, near $r=0$ the metric takes the form 
\begin{equation}
ds^2 \simeq -\frac{4Q^2}{\asr}(dt - \frac{a}{\sqrt{1-
\frac{a^2\alpha^2}{2}}}d\varphi)^2.
                             \label{eq:4138}
\end{equation}
If one now changes coordinates to  $x = r\cos\varphi- {a}(1-
\frac{a^2\alpha^2}{2})^{-\frac12}\sin\varphi$ and 
$y = r\sin\varphi + {a}(1-
\frac{a^2\alpha^2}{2})^{-\frac12}\cos\varphi$ one finds that at
$r\rightarrow0$ the metric is given 
by
\begin{equation}
ds^2 \simeq -\frac{4Q^2}{\asr}\left\lbrack dt -
\frac{\sqrt{1-\frac{a^2\alpha^2}{2}}}{a}(xdy-ydx)\right\rbrack^2.
                             \label{eq:4139}
\end{equation}
Now, $x^2+y^2= \frac{a^2}{1-\frac{a^2\alpha^2}{2}} + r^2 \simeq 
\frac{a^2}{1-\frac{a^2\alpha^2}{2}}$. Thus the singularity at $r=0$, 
like the Kerr-Newman singularity, has a ring structure.  However, 
unlike Kerr-Newman,  one cannot penetrate to the inside of the 
singularity. 

From (\ref{eq:4004}) one can see that there is an infinite redshift surface 
given by the zero of the coefficient in front of $dt^2$. It is also a 
quartic equation and one can easily find $r_{\rm rs}$. It is always outside 
the event horizon, unless $a=0$ in which case it coincides 
with the event horizon. We also note that if there is no rotation 
$a=0$ then the Penrose diagram in figure  4.2 is identical. 
However there are no CTCs and 
the singularity looses the ring structure. 
\vskip 0.3cm

(ii) $a^2\alpha^2 = \frac23 - \frac{128}{81}
\frac{Q^6}{M^4(1-\frac12a^2\alpha^2)^3}$

The extreme case is given when $Q$ is connected to $M$ and $a$ 
through the relation,
\begin{equation}
Q^6= \frac{27}{64} M^2 {\left(1-\frac{3}{2}a^2\alpha^2\right)
\left(1-\frac{1}{2}a^2\alpha^2\right)^3},
                            \label{eq:4201}
\end{equation}
which can also be put in the form 
$a^2\alpha^2 = \frac23 - \frac{128}{81}
\frac{Q^6}{M^4(1-\frac12a^2\alpha^2)^3}$ as above. 
In figure  4.1
we have drawn the line which gives the values of $Q$ and $a$ 
(in suitable $M$ units) 
compatible with this case. The event and Cauchy horizons join together 
in one single horizon $r_+$ given by
\begin{equation}
r_+ = \frac{4Q^2}{3M\alpha(1-\frac{a^2\alpha^2}{2})}
                             \label{eq:4202}
\end{equation}
The function $\Delta$ is now, 
\begin{equation}
\Delta = \frac{\alpha^2(r-{r_+})^2(r^2+2r_+r+3{r_+}^2)}{r^2}
                             \label{eq:4203}
\end{equation}
so the metric (\ref{eq:4101}) turns to 
\begin{eqnarray}
&ds^2=-
\frac{\alpha^2(r-{r_+})^2(r^2+2r_+r+3{r_+}^2)}{r^2}(\gamma dt
-\frac{\omega}{\alpha^2}d\varphi)^2 + 
\frac{r^2dr^2}{\alpha^2(r-{r_+})^2(r^2+2r_+r+3{r_+}^2)}+&
\nonumber \\
&+r^2(\gamma d\varphi - \omega dt)^2 + \asr dz^2,&
                             \label{eq:4204}
\end{eqnarray}
where $\gamma$ and $\omega$ are defined in  (\ref{eq:4105}) and 
(\ref{eq:4106}) respectively. 
There are no Kruskal coordinates. To draw the Penrose diagram we resort 
first to the double null coordinates $u$ and $v$, 
\begin{equation}
u=\alpha(\frac t\gamma -  r_*) 
\quad {\rm and} \quad v=\alpha(\frac t\gamma + r_*)
                            \label{eq:4205}
\end{equation}
where $r_*$ is the tortoise coordinate given by 
\begin{eqnarray}
& r_* = \frac{2}{9\alpha r_+} \ln \left\lbrack \frac{\alpha(r-r_+)}
{b^{\frac13}}\right\rbrack -\frac{1}{6\alpha(r-r_+)} 
-\frac{1}{9\alpha r_+}\ln \left\lbrack \frac{\alpha^2(r^2+2r_+r+3{r_+}^2)}
{b^{\frac23}} \right\rbrack + &
\nonumber \\
& + \frac{7}{18\sqrt2\alpha r_+}\arctan\frac{r+r_+}{\sqrt2 r_+}.&
                            \label{eq:4206}
\end{eqnarray}
Defining the new angular coordinate as before $\overline{\varphi} =
\gamma \varphi -\omega t$,  the metric (\ref{eq:4204}) is now
\begin{eqnarray}
&ds^2=-
\frac{\alpha^2(r-{r_+})^2(r^2+2r_+r+3{r_+}^2)}{r^2} \frac{dt^2}{\gamma^2}+
\frac{r^2dr^2}{\alpha^2(r-{r_+})^2(r^2+2r_+r+3{r_+}^2)}+&
\nonumber \\
&+\frac{\Delta\omega}{\alpha^2\gamma^2} 2dt d\overline{\varphi} + 
(r^2 - \Delta \frac{\omega^2}{\alpha^4\gamma^2})d\overline{\varphi}^2+
\asr dz^2. &
                             \label{eq:4207}
\end{eqnarray}
Now define the Penrose coordinates \cite{carter2,bhtz} $\psi$ and 
$\xi$ via the relations
\begin{equation}
u=\tan\frac12 (\psi-\xi) \quad {\rm and} \quad v=\tan\frac12(\psi+\xi)
                             \label{eq:4208}
\end{equation}
Then the metric (\ref{eq:4207}) turns into
\begin{eqnarray}
&ds^2=-
\frac{(r-{r_+})^2(r^2+2r_+r+3{r_+}^2)}{r^2} 
\frac{d\psi^2-d\xi^2}{(\cos \psi + \cos \xi)^2} +&
\nonumber \\
&+\frac{\Delta\omega}{\alpha^2\gamma^2} 2dt d\overline{\varphi} + 
(r^2 - \Delta \frac{\omega^2}{\alpha^4\gamma^2})d\overline{\varphi}^2+
\asr dz^2, &
                             \label{eq:4209}
\end{eqnarray}
where $t$ is given implicitly in terms of $\psi$ and $\xi$. 
From the defining equations (\ref{eq:4205}) and (\ref{eq:4208})
we have 
\begin{equation}
\frac{\sin \xi}{\cos \psi + \cos \xi} = \alpha r_*
                             \label{eq:4210}
\end{equation}
Then one can draw the Penrose diagram (see figure  4.3). The lines 
$r=r_+$ are given by the equation $\psi =\pm \xi + n\pi$ with $n$ any integer, 
and therefore are lines at $45^0$.  The lines $r=0$ and $r=\infty$ are 
timelike lines given by an equation of the form 
$\frac{\sin \xi}{\cos \psi + \cos \xi} = {\rm constant}$, where the 
constant is 
easily found from $r_*$. These are not straight vertical lines. However 
by a further coordinate transformation it is possible to straighten 
them out as it is shown in the figure. 
The metric   (\ref{eq:4209}) is regular at $r=r_+$ because the zeros 
of the denominator and numerator cancel each other.

The CTC radius is defined in 
(\ref{eq:4135}), where now we take the values for $Q$ and $a$ 
appropriate for the extreme case. This line is also shown in the 
diagram. There is 
also an infinite redshift line, $r_{\rm rs}$. If $a=0$ there are 
no CTCs and $r_{\rm rs}$ joins the event horizon.
\vskip 0.3cm

(iii)  $\frac23 - \frac{128}{81}
\frac{Q^6}{M^4(1-\frac12a^2\alpha^2)^3} < a^2\alpha^2<\frac23$

In this case there are no roots for $\Delta$ as defined in 
(\ref{eq:4102}). Therefore there are no horizons. The singularity 
is timelike and naked. Infinity is also timelike. 
There is an infinite redshift surface if the following inequality is 
satisfied
\begin{equation}
{Q}^6 \leq \frac{27}{64}(1-\frac12a^2\alpha^2)^4M^4.
                               \label{eq:4301}
\end{equation}
There are CTCs  if $a\neq0$. The Penrose diagram is sketched in 
figure  4.4. 
\vskip 0.3cm

(iv) $a^2\alpha^2=\frac23$ 

For $a^2\alpha^2=\frac23$ the form of the metric (\ref{eq:4101}) is 
not valid, one has to go back to the form  (\ref{eq:4004}).
The Kretschmann scalar is a constant. 
However, since the first Betti number of the manifold is one, 
there is a topological singularity at $r=0$, which is a 
null surface. There is an infinite redshift surface and a closed 
timelike radius. 
The Penrose diagram is sketched in 
figure  4.5. 
\vskip 0.3cm

(v) $\frac23<a^2\alpha^2 \leq1$

Here $\Delta$ has one root, and therefore there is one horizon. 
Thus, it represents a black 
hole. The singularity at $r=0$ is polynomial 
and spacelike. The CTC radius is outside the event horizon. If $Q$ 
is sufficiently small there is an ergosphere, see figure 4.6.

\vskip 0.5cm

{\bf IV.2} \quad $r\leq0$ (or $M<0$)

In order to be complete we continue to analyze the solution found in 
(\ref{eq:4004}). The solutions displayed below are rather exotic, 
although not without interest.

When we say $r\leq0$ it amounts to do $r\rightarrow -r$ in the 
metric given in equation (\ref{eq:4004}), or equivalently 
$M\rightarrow-M$. Thus the following solution is a negative mass 
solution. The metric is then, 
\begin{eqnarray}
&ds^{2} = -\left(\asr +\frac{4M(1-\frac{a^2\alpha^2}{2})}{\alr} + 
\frac{4Q^2}{\asr}\right) dt^2 + &
\nonumber \\
&-\frac{4aM\sqrt{1-\frac{a^2\alpha^2}{2}}}{\alr}\left(1+
\frac{Q^2}{M(1-\frac{a^2\alpha^2}{2})\alr}\right) 2dt d\varphi +&
\nonumber \\
&+ \left(\asr +\frac{4M(1-\frac{3}{2}a^2\alpha^2)}{\alr} + 
\frac{4Q^2}{\asr} 
\frac{(1-\frac{3}{2}a^2\alpha^2)}{(1-\frac12{a^2\alpha^2})}
\right)^{-1} dr^2 +&
\nonumber \\
&+ \left[r^2 - \frac{4Ma^2}{\alr}\left(1+
\frac{2Q^2}{(1-\frac{a^2\alpha^2}{2})M\alr}\right)\right] d\varphi^2 + 
\asr dz^2,&
                             \label{eq:4401}
\end{eqnarray}
where we have done $J\rightarrow -J$ 
(or $a\rightarrow-a$) since this does not change the character 
of the metric. Now, the function $\Delta$ is 
\begin{equation}
\Delta= \asr +\frac{b}{\alr}+\frac{c^2}{\asr},
                             \label{eq:4402}
\end{equation}
with $b=4M\left(1-\frac{3}{2}a^2\alpha^2\right)$ and 
$c^2= 4Q^2\left( \frac{1-\frac{3}{2}a^2\alpha^2}
{1-\frac12{a^2\alpha^2}}\right)$ as in equations 
(\ref{eq:4103}) and (\ref{eq:4104}). However, the 
parameter $M$ still positive, represents now negative mass.
Note then that whether $\Delta$ has roots or not depends only 
on the value of $a$. Recall that $0\leq a^2\alpha^2\leq 1$. Thus, 
looking at the $g_{00}$ term of (\ref{eq:4401}) one sees that 
there are no infinite redshift surfaces. 
There are three distinct cases. 
\vskip 0.3cm

(i) $0< a^2\alpha^2< \frac{2}{3}$

Here $\Delta>0$ always, therefore there are no roots, no horizons. 
There is a timelike singularity at $r=0$. At $r\rightarrow\infty$ 
spacetime is anti-de Sitter. There is a $r_{\rm CTC}$. 
Supressing the infinite redshift surface $r_{\rm rs}$, figure 4.4  
serves as a representation of the Penrose diagram.  
If $a=0$ there are no CTCs.
\vskip 0.3cm

(ii) $a^2\alpha^2 = \frac{2}{3}$

There is a topological (non-polynomial) null singularity at $r=0$. 
The singularity is topological because the first Betti number of the 
manifold is one. See figure  4.5.  If $a=0$ there are no CTCs.
See figure  4.5 (with $r_{\rm rs}$ supressed) 
for a representation of the Penrose diagram.
\vskip 0.3cm

(iii) $\frac{2}{3}< a^2\alpha^2< 1$

In this case there is one root in $\Delta$ for positive $r$, therefore 
there is one horizon located at 
\begin{equation}
r_h = \left\vert b\right\vert^\frac13\frac{\sqrt{\sigma} + 
\sqrt{2\sqrt{\sigma^2+4 q^2}-\sigma}}{2\alpha},
                             \label{eq:4403}
\end{equation}
where, 
\begin{equation}
\sigma = \left( \frac12 + 
\frac12 \sqrt{1+4\left(\frac{4 q^2}{3}\right)^3}\right)
^{\frac13} +
 \left( \frac12 - \frac12 \sqrt{1-4\left(\frac{4q^2}{3}\right)^3}\right)
^{\frac13},
                             \label{eq:4404}
\end{equation} 
and here $q = \frac{\left\vert c^2\right\vert}{\left\vert b
\right\vert^\frac43}$. There are also CTCs. It is easy to check 
that $r_{\rm CTC}>r_h$. See figure  4.6 (with $r_{\rm rs}$ supressed) 
for the Penrose diagram. For $a=0$  there are no CTCs.

\vspace*{1.cm}
\addtocounter{chapter}{1}
\setcounter{equation}{0}
\setcounter{figure}{0}
\vspace{10pt}

{\bf V. Causal Structure of the Uncharged Rotating Black String Spacetime}

This case has been proposed in \cite{lemos1,lemos2}. 
Since the solution and its 
causal structure are totally different in character from the charged 
solution we  present it here in detail.

For $Q=0$ the metric  (\ref{eq:4004}) simplifies to 
\begin{eqnarray}
&ds^{2} = -\left(\asr -\frac{4M(1-\frac{a^2\alpha^2}{2})}{\alr} 
\right) dt^2 + &
\nonumber \\
&-\frac{4aM\sqrt{1-\frac{a^2\alpha^2}{2}}}{\alr} 2dt d\varphi +&
\nonumber \\
&+ \left(\asr -\frac{4M(1-\frac{3}{2}a^2\alpha^2)}{\alr} 
\right)^{-1} dr^2 +&
\nonumber \\
&+ \left( r^2 + \frac{4Ma^2}{\alr}\right) d\varphi^2 + 
\asr dz^2.&
                             \label{eq:51}
\end{eqnarray}
It can also be put in a form like (\ref{eq:4101})
\begin{eqnarray}
&ds^{2} = -(\asr -\frac{b}{\alr} )
\left( \gamma dt - \frac{\omega}{\alpha^2} d\varphi\right)^2 + &
\nonumber \\
&+r^2 \left(\gamma d\varphi - \omega dt\right)^2 +
\frac{dr^2}{\asr -\frac{b}{\alr}} +
\asr dz^2. &
                             \label{eq:52}
\end{eqnarray}
and now we have 
\begin{equation}
\Delta = \asr -\frac{b}{\alr}
                             \label{eq:53}
\end{equation}
with $b= 4M\left(1-\frac{3}{2}a^2\alpha^2\right)$ as in (\ref{eq:4103}) and 
$\gamma$ and $\omega$ given as in (\ref{eq:4105}) and (\ref{eq:4106})
respectively. 
The solution changes character depending on whether $r\leq0$ or 
$r\geq0$ as in the charged case. We study first $r\geq0$ which 
contains the most interesting rotating black hole solution.
\vskip 0.5cm

{\bf V.1}\quad $r\geq0$

From (\ref{eq:51}) we see that to have horizons one needs 
\begin{equation}
0\leq a^2\alpha^2< \frac23
                             \label{eq:54}
\end{equation}
One has therefore three distinct cases which depend on the value 
of $a$.
\vskip 0.3cm

(i) $0< a^2\alpha^2< \frac{2}{3}$

In this case $\Delta$ has one root, which locates the horizon at the 
radius
\begin{equation}
r_+ = \frac{b^\frac13}{\alpha} = 
\frac{\left\lbrack 4M(1-\frac32 a^2\alpha^2)\right\rbrack^{\frac13}}{\alpha}
                             \label{eq:55}
\end{equation}

To find the Kruskal coordinates we first define the tortoise coordinate 
$r_* = \int \frac{dr}{\Delta}$. Using (\ref{eq:53})  we find
\begin{equation}
r_* = \frac{1}{\alpha^2 r_+} 
\left\lbrack \frac16 \ln\frac{(r-r_+)^2}{r^2 + r_+r+{r_+}^2} 
+\frac{1}{\sqrt3} \arctan\frac{2r+r_+}{\sqrt3 r_+}\right\rbrack 
                             \label{eq:56}
\end{equation}
Then the Kruskal coordinates are for $r\geq r_+$,
\begin{eqnarray}
& U = - {\rm e}^{-\frac32 \alpha^2 
r_+(\frac{t}{\gamma}-r_*)} = 
-(r-r_+)^\frac12 \left\lbrack \frac
{{\rm{e}}^{\sqrt3\arctan\frac{2r+r_+}{\sqrt3 r_+}}}
{(r^2+r_+r+{r_+}^2)^\frac12}
\right\rbrack^\frac12
{\rm e}^
{-\frac32  \frac{\alpha^2 r_+t}{\gamma}}\, ,
&
\nonumber \\
& V =  {\rm e}^{-\frac32 \alpha^2 
r_+(\frac{t}{\gamma}+r_*)} = 
(r-r_+)^\frac12 \left\lbrack \frac
{\rm{e}^{\sqrt3\arctan\frac{2r+r_+}{\sqrt3 r_+}}}
{(r^2+r_+r+{r_+}^2)^\frac12}
\right\rbrack^\frac12
{\rm e}^
{\frac32  \frac{\alpha^2 r_+ t}{\gamma}}\, .&
                             \label{eq:57}
\end{eqnarray}
And for $0<r<r_+$ we choose the following coordinates,  
\begin{eqnarray}
& U =  (r_+-r)^\frac12 \left\lbrack \frac
{\rm{e}^{\sqrt3\arctan\frac{2r+r_+}{\sqrt3 r_+}}}
{(r^2+r_+r+{r_+}^2)^\frac12}
\right\rbrack^\frac12
{\rm e}^
{-\frac32  \frac{\alpha^2 r_+t}{\gamma}}\, ,
&
\nonumber \\
& V =  (r_+-r)^\frac12 \left\lbrack \frac
{\rm{e}^{\sqrt3\arctan\frac{2r+r_+}{\sqrt3 r_+}}}
{(r^2+r_+r+{r_+}^2)^\frac12}
\right\rbrack^\frac12
{\rm e}^
{\frac32  \frac{\alpha^2 r_+ t}{\gamma}},&
                             \label{eq:58}
\end{eqnarray}

\noindent Then defining the usual new angular coordinate 
$\overline{\varphi} = \gamma d\varphi -\omega dt$ one finds that 
metric (\ref{eq:51}) is written in Kruskal coordinates as 
\begin{eqnarray}
&ds^2 = -\frac{4}{9\alpha^2 {r_+}^2} 
\frac{(r^2+r_+r+{r_+}^2)^\frac32}
{r {\rm e}^{\sqrt3 \arctan\frac{2r+r_+}{\sqrt3 r_+}}}
dUdV + &
\nonumber \\
&+\frac{2}{3r_+}
\frac
{(r^2+r_+r+{r_+}^2)^\frac32}
{r {\rm e}^{\sqrt3 \arctan\frac{2r+r_+}{\sqrt3 r_+}}}
\frac{a}{\sqrt{1-\frac{a^2\alpha^2}{2}}}
(VdU-UdV)d\overline{\varphi} + &
\nonumber \\
&+(r^2 - \Delta \frac{a^2}{1-\frac{a^2\alpha^2}{2}})d\overline{\varphi}^2 
+\asr dz^2.&
                             \label{eq:59}
\end{eqnarray}
The metric is regular at $r=r_+$. To draw the Kruskal diagram it is 
a simple matter. One has only to take the  product
$UV$ appropriately from equations  (\ref{eq:57})
and  (\ref{eq:58}). Then one has the usual hyperbolas in the 
Kruskal diagram. To revert to the Penrose diagram one takes 
as usual the arctangents of $U$ and $V$. The diagram is represented
in figure  5.1.

There are no CTCs. The infinite redshift surface $r_{\rm rs}$ is given when
$g_{00}=0$ in  (\ref{eq:51}), i.e., $r_{\rm rs} = \frac1\alpha
\left\lbrack 4M(1-\frac32 a^2\alpha^2 )\right\rbrack^\frac13 $. If the
black hole has no rotation, $a=0$, then the Penrose diagram looks the
same. However, in this case the infinite redshift surface coincides 
with the event horizon. 
\vskip 0.3cm

(ii) $a^2\alpha^2 = \frac{2}{3}$

In this case $b=0$ The Kretschmann scalar is constant and there is no
polynomial singularity .  However there is a topological singularity at
$r=0$. The singularity is a null line. There is an infinite redshift
surface at $r_{\rm rs} = (\frac{2Ma^2}{\alpha})^\frac13$.  This can be
considered the extremal uncharged black hole. See figure  5.2.

\vskip 0.3cm

(iii) $ \frac{2}{3}< a^2\alpha^2 <1$

In this case $\Delta$ is positive for all $r$. There are no horizons. 
There is still an infinite redshift surface. The singularity at $r=0$ 
is timelike. At $r\rightarrow\infty$ spacetime is anti-de Sitter. 
The Penrose diagram is represented in figure 5.3.

\vskip 0.5cm

{\bf V.2}\quad $r\leq0$

When we do $r\rightarrow-r$ the metric  (\ref{eq:51}) turns into 
(\ref{eq:4401}) with $c^2=0$, i.e., $Q=0$. For $r\geq0$ we have just 
seen that the character of the solution is totally different depending 
on whether it has charge or not. For $r\leq0$ the character of the 
solution doesnot change at all. Thus there are the same three cases 
as in section IV.2 and the Penrose diagrams are simply the same.
\vspace*{1.cm}
\addtocounter{chapter}{1}
\setcounter{equation}{0}
\setcounter{figure}{0}
\vspace{10pt}

{\bf VI. Causal Structure of Other Solutions}

Now, when solving equation (4.4) for $\gamma$ and $\omega$ we obtained 
quadratic equations. The solutions we have discussed so far in sections 
IV and V are all solutions with positive sign in front of the 
square root of the discriminant, see equation (\ref{eq:36}).
We now discuss solutions with 
negative sign, given by (\ref{eq:37}). 
This amounts to do $\Omega\rightarrow-\Omega$ in (\ref{eq:310}). Then with 
the definitions (\ref{eq:4001}) (\ref{eq:4002}) we obtain the following 
metric,
\begin{eqnarray}
&ds^{2} = -\left(\asr -\frac{2Ma^2\alpha^2}{\alr} + 
\frac{4\overline{Q}^2a^2\alpha^2}{\asr}\right) dt^2 + &
\nonumber \\
&-\frac{4aM\sqrt{1-\frac{a^2\alpha^2}{2}}}{\alr}\left(1-
\frac{2\overline{Q}^2}{M\alr}\right) 2dt d\varphi +&
\nonumber \\
&+ \left(\asr + \frac{8M(1-\frac{3}{4}a^2\alpha^2)}{\alr} - 
\frac{16\overline{Q}^2(1-\frac34 a^2\alpha^2)}{\asr} 
\right)^{-1} dr^2 +&
\nonumber \\
&+ \left[r^2 + \frac{8M(1-\frac{a^2\alpha^2}{2})}{\alpha^3 r}\left(1-
\frac{2\overline{Q}^2}{M\alr}\right)\right] 
d\varphi^2 + \asr dz^2,&
                             \label{eq:61}
\end{eqnarray}
where we have defined
\begin{equation}
\overline{Q}^2=\frac{Q^2}{a^2\alpha^2}.
                             \label{eq:62}
\end{equation}
Then to study the causal structure one can subdivide it again in $r\geq0$ 
and $r\leq0$. 
\vskip 0.5cm

{\bf VI.1}\quad $r\geq0$

These are the positive mass solutions. We have two distinct cases. Indeed, 
if $\overline{Q}\neq0$ there is one horizon only, $r_h$. There are 
CTCs and $r_{\rm CTC}>r_h$ always.  There is an infinite redshift surface 
for certain values of $a$ and $Q$. If $\frac{512\overline{Q}^6}{a^2\alpha^2}
>27 M^4$ there are no infinite redshift surfaces 
(see figure  4.6 for the Penrose diagram). 
On the other hand 
if $\overline{Q}=0$ then there are no horizons, the singularity is naked.
There is an infinite redshift surface and there are no CTCs. The Penrose 
diagram looks like figure  4.4. 

\vskip 0.5cm

{\bf VI.2}\quad $r\leq0$

In this case the metric reads,
\begin{eqnarray}
&ds^{2} = -\left(\asr +\frac{2Ma^2\alpha^2}{\alr} + 
\frac{4\overline{Q}^2a^2\alpha^2}{\asr}\right) dt^2 + &
\nonumber \\
&-\frac{4aM\sqrt{1-\frac{a^2\alpha^2}{2}}}{\alr}\left(1+
\frac{2\overline{Q}^2}{M\alr}\right) 2dt d\varphi +&
\nonumber \\
&+ \left(\asr - \frac{8M(1-\frac{3}{4}a^2\alpha^2)}{\alr} - 
\frac{16\overline{Q}^2(1-\frac34 a^2\alpha^2)}{\asr} 
\right)^{-1} dr^2 +&
\nonumber \\
&+ \left[r^2 - \frac{8M(1-\frac{a^2\alpha^2}{2})}{\alpha^3 r}\left(1+
\frac{2\overline{Q}^2}{M\alr}\right)\right] 
d\varphi^2 + \asr dz^2,&
                             \label{eq:63}
\end{eqnarray}
Now, this is a solution with negative mass. $M$ represents negative 
mass. From the term in $dr^2$ we see that there is a 
horizon always. From the $g_{00}$ component we see that 
there are no infinite redshift surfaces. 
There are CTCs and one can check that $r_{\rm CTC}>r_H$ always 
(see figure  4.6).

\vspace*{1.cm}
\addtocounter{chapter}{1}
\setcounter{equation}{0}
\setcounter{figure}{0}
\vspace{10pt}

{\bf  VII. Geodesic structure}

We study the geodesic equations with emphasis in the
null geodesics. We find the equations and their first integrals.
We study the effective potential and show the possible turning
points.  We comment on the timelike
geodesics.
The results, very interesting, underline the similarities
and non-similarities with the Kerr and Schwarzschild black hole.

 As we have seen in the previous sections, there are several particular cases
whose geodesic structure should be studied separately. However, we are
going to fix attention on the two most relavant cases. Namely, the black
string with two horizons considered in section IV.1(i) and the extremal
black string case of section IV.1(ii).

The situation we are interested here is given in equation (\ref{eq:4108}). 
Then, the horizons $r=r_{+}$  (event horizon) and $r=r_{-}$
(Cauchy horizon) are given respectively by equations
(\ref{eq:4112}) and (\ref{eq:4113}). There is also an infinite redshift
surface outside the horizon at $r=r_{\rm rs}$ as shown in figure 4.2.
For $r < r_{\rm rs}$ all observers with fixed $r$ and $z$  must orbit the black
hole in the direction it rotates.

In order to be more specific we must study frame dragging in 
metric (\ref{eq:4004}). Thus, the angular velocity $\Omega = d\varphi/dt$ of
stationary observers is constrained by
\begin{eqnarray}
& &\Omega_{min} = \omega -\sqrt{\omega^2 -g_{tt}/g_{\varphi\varphi}}<\Omega
<\Omega_{max} = \omega + \sqrt{\omega^2 -g_{tt}/g_{\varphi\varphi}}\,
\nonumber \\
& & \omega = \frac{4aM\sqrt{1-\frac{a^2\alpha^2}{2}}\left(1-
\frac{Q^2}{M(1-\frac{a^2\alpha^2}{2})\alr}\right)} {\alpha r^3 +
4Ma^2\left(1-\frac{Q^2}{(1-\frac{a^2\alpha^2}{2})M\alr}\right)}
                             \label{eq:701}
\end{eqnarray}
From (\ref{eq:701}) we see that $\Omega_{min}$ is null at $r= r_{\rm rs}$ 
and is positive for $r<r_{rs}$, which means that there is no static 
observers inside such a surface.
It is also worth to note that for
$r\longrightarrow \infty$, $\omega$ falls
as $r^{-3}$ just like in the case of the Kerr metric and
in constrast with BTZ \cite{btz} black hole where $\omega =
J/2r^2$.
Let us now turn our attention to the geodesic equations.

In the metric (\ref{eq:4004}) there are three Killing vectors: one of them
$\varepsilon^{\mu} \equiv \xi_{t}^{\mu} 
= (\frac{\partial}{\partial t})^{\mu}$ is 
timelike, associated to the time
translation invariance, and the two others $\phi^{\mu} \equiv
\xi_{\varphi}^{\mu} = 
(\frac{\partial}{\partial\varphi})^{\mu}$ 
and $\zeta^{\mu} \equiv \xi_{z}^{\mu} = 
=(\frac{\partial}{\partial z})^{\mu}$ are both
spacelike associated to the invariance of the metric under rotations around
the simmetry axis and $z$-translations, respectively 
(we are considering the standard cylindrically symmetric model, 
$G_2= R\times U(1)$). Then, considering the
geodesic motion in such a spacetime, the (three) constants related to these
Killing vectors are:
\begin{equation}
 E = - g_{\mu\nu}\varepsilon^{\mu}u^{\nu}\, ; \hspace{1cm}
 L =  g_{\mu\nu}\phi^{\mu}u^{\nu}\, ; \hspace{1cm}
 P =  g_{\mu\nu}\zeta^{\mu}u^{\nu}\, ,\label{eq:702}
\end{equation}
where $u^\mu = \dspst{\frac{dx^\mu}{d\tau}}$ with $\tau$ being a
(affine) parameter on the geodesic curve.
 As $u^\mu$ is the tangent  vector to the curve it may be
normalized as
\begin{equation}
u^{\mu}u_{\mu} = - \epsilon^2 \, ,  \label{eq:703}
\end{equation}
with $\epsilon^2 = 1 (0)$ for timelike (null) geodesics.

Using (\ref{eq:702}), (\ref{eq:703}) and metric (\ref{eq:4004}) we may write
the geodesic equations in the form
\begin{eqnarray}
\dot{t}& =& \frac{1}{Br^2}\left(E\alpha^2r^4+4aMc_{0}\alr-\frac{4aQ^2c_{0}}
{1-\frac{a^2\alpha^2}{2}}\right)\, \nonumber\\
\dot{\varphi} &=&\frac{1}{Br^2}\left(L\alpha^4r^4+4Mc_{0}
\sqrt{1-\frac{\alpha^2a^2}{2}}\alr-\frac{4Q^2c_{0}}
{\sqrt{1-\frac{\alpha^2a^2}{2}}}\right)\, , \nonumber \\
\dot{z} &=& P/\asr \, , \nonumber\\
\alpha^4 r^4 \dot{r}^2 & = & -\left(\epsilon^2\asr +P^2 +
\frac{\alpha^2c_{0}^2}{1-\frac{3\alpha^2a^2}{2}}\right)B
+ \alpha^4r^4\frac{c_{1}^2}{1-\frac{3\alpha^2a^2}{2}}\, ,
                   \label{eq:704}
\end{eqnarray}
where $\dot{t}=\frac{dt}{d\tau}$, etc.
In the above equations we have introduced the definitions
\begin{eqnarray}
c_{0}&=&Ea -L\sqrt{1-\frac{\alpha^2a^2}{2}} \, , \label{eq:705} \\
c_{1}&=&E\sqrt{1-\frac{\alpha^2a^2}{2}}-\alpha^2La\, , \label{eq:706}\\
B&=& \alpha^4 r^4 -4M\left(1-\frac{3a^2\alpha^2}{2}\right)\alr + 4Q^2
\frac{\left(1-\frac{3a^2\alpha^2}{2}\right)}{(1-\frac{a^2\alpha^2}{2})} \, .
                       \label{eq:707}
\end{eqnarray}
It is worth to note that since the spacetime is not asymptotically flat the
constants $E$ and $L$ cannot be interpreted as the local energy and angular
momentum at infinity.

In order to study the behaviour of geodesic lines in the black string 
spacetime, the important equation to analyze is the one that dictates 
the behaviour along the radial coordinate.  It is not easy to give a 
complete description of the motion of the $r$ coordinate, because the 
governing function, i. e., the right hand side of (\ref{eq:704}), 
is in fact a polynomial of sixth degree and the equation cannot be 
exactly integrated. Neverthless, it is possible after some algebraic 
and numerical manipulations to reach interesting conclusions. 

To simplify discussions let us study first the static charged solution 
where $a^2=0$ and then the  general case with rotation. 

\vskip 0.5cm

{\bf VII.1 Static Case}

 In the static case $J = 0$, or equivalently $a=0$, and the equation for
$\dot{r}$ may be used to find the main properties of timelike 
and null geodesics. Then by putting $a =0$ in (\ref{eq:704}) we get
\begin{eqnarray}
\dot{t}& = & \frac{E\asr}{B}\, , \hspace{1cm}
\dot{\varphi} = \frac{L}{r^2}\, ,  \hspace{1cm}
\dot{z} =  \frac{P}{\asr} \, , \label{eq:708}\\
\dot{r}^2 & = & E^2 - V^2_{eff}\, , \hspace{0.1cm}
V^2_{eff}=\left(\epsilon^2\asr +P^2 +\alpha^2 L^2\right)
\frac{B}{\alpha^4r^4}\, , \label{eq:709}
\end{eqnarray}
where
\begin{equation}
B = \alpha^4r^4 - 4M\alr +4Q^2\, . \label{eq:710}
\end{equation}
Notice that the equation for the radial coordinate has the same polynomial
form as (\ref{eq:704}). In the static
charged black string case, condition (\ref{eq:4108}) reduces to
\begin{equation}
 Q^6\leq \frac{27}{64}M^4 \, . \label{eq:711}
\end{equation}
Let us the consider each case of (\ref{eq:711}) in turn.

\vskip 0.3cm

(a) $ Q^6<  \frac{27}{64}M^4$

Then, $B$ has two roots $r_{\pm}$ give
respectively by equations (\ref{eq:4112})--(\ref{eq:4115}) with $a=0$.

\begin{equation}
r_+ =(4M)^\frac13\frac{\sqrt{s}+\sqrt{2\sqrt{s^2-
      Q^2\left(\frac{2}{M}\right)^\frac43}-s}}{2\alpha}
                             \label{eq:712}
\end{equation}
and
\begin{equation}
r_- =(4M)^\frac13 \frac{\sqrt{s} - \sqrt{2\sqrt{s^2-
  Q^2\left(\frac{2}{M}\right)^\frac43}-s}}{2\alpha}
                             \label{eq:713}
\end{equation}
where,
\begin{equation}
s = \left( \frac12 + \frac12 \sqrt{1-\frac{64Q^6}{27M^4}}\right)^{\frac13}+
 \left( \frac12 - \frac12 \sqrt{1-\frac{64Q^6}{27M^4}}\right)^{\frac13},
                             \label{eq:714}
\end{equation}
From the explicit form of the effective potential (\ref{eq:709})
we see that there are two turning points, where $\dot{r}^2=0$, $r_{2}$ and
$r_{1}$, with $r_{2}>r_{+}$ and $r_{1}<r_{-}$, for any finite values of the
parameters of motion $E$, $L$, and $P$.  The allowed region
for geodesic particles is between $r_{1}$ and $r_{2}$ where the right hand
side of (\ref{eq:709}) is a positive number. For null geodesics with
$E^2 -\alpha^2L^2 -P^2 \geq 0$ the turning point $r_{2}$ is at infinity.
Notice also that for
motion with $E=0$, the two turning points coincide with the horizons and
such geodesic particles must be confined to this region.

Thus,  due to the simplicty of the function $V_{eff}$,  
it is easy to see from (\ref{eq:709}) and (\ref{eq:710}) how the radial
coordinate varies during the geodesic motion, (even without integrating 
such equation exactly). 
For null geodesics the effective potencial is given by
\begin{equation}
V^2_{eff}=  (P^2 +\alpha^2 L^2)\left(1 - \frac{4M}{\alpha^3r^3}+
\frac{4Q^2}{\alpha^4r^4}\right)\, . \label{eq:715}
\end{equation}
Then, from (\ref{eq:709}) and (\ref{eq:715}) we can draw the following
conclusions.

(i) If $E^2 -\alpha^2L^2 - P^2 > 0$ and $\alpha^2L^2 +P^2\neq 0$
null particles produced near the horizon
may escape to $r=+\infty$, while particles coming in from infinity reach a
minimum distance from the black string where they are then scattered by
the potential barrier and spiral back to infinity.

(ii) Null geodesic particles with
$E^2 - \alpha^2L^2 - P^2 < 0$ produced  near the horizon
reach a maximum distance from the black string at $r=r_{2}$
and are scattered to $r=r_{1}$. Then, the
radial coordinate $r$ starts to increase reaching $r_{2}$ again, and so on.
The geodesic oscilates indefinitely between $r_{1}$ and $r_{2}$
crossing the horizons an infinite number of times.

(iii) Just radial null geodesics can reach the singularity $r=0$.
To see it notice that $V_{eff}=0$ ($P^2=0$, $L^2=0$) and then
the radial equation yields
\begin{equation}
r= r_{0} \pm E\tau\, , \label{eq:716}
\end{equation}
where $r_{0}$ is an integration constant. Such relation shows that this kind
of line extends either from infinity to the singularity or from the
singularity to infinity.

(iv) If $E^2 =\alpha^2 L^2 +P^2$ the equation for $\frac{dr}{d\tau}$
may be integrated exactly producing
\begin{eqnarray}
\pm 15\sqrt{\alpha M(P^2+\alpha^2L^2)}\,\tau+{\rm const.}
&=&\left(3r^2+\frac{4Q^2}
{\alpha M}r +\frac{8Q^2}{\alpha^2M^2}\right)\sqrt{r-\frac{Q^2}{\alpha M}}\, ,
\nonumber \\
\mbox{or,}\hspace*{1cm}
r&=&\frac{Q^2}{\alpha M}+\frac{ME^2}{\alpha^3L^2}(\varphi-\varphi_{0})^2\, ,
\label{eq:717}
\end{eqnarray}
where $\varphi_{0}$ is a constant. Then we can see that in such a case
null geodesics are spirals that start at $r_{1}= \frac{Q^2}{\alpha M}$ and
may reach infinite radial distances. 
They can also start at $r=+\infty$ spiralling to $r_{1}$ and
then getting back to infinity. 
In the same way, some conclusions may be found for timelike geodesics.

(v) From the asymptotic form of $V_{eff}$ it is easy to see that
timelike geodesics cannot reach the singularity $r=0$ nor escape to
$r\rightarrow +\infty$. The trajectory of timelike geodesic particles are
bounded between $r_{1}$ and $r_{2}$. The behaviour is the same as described
in (ii) for null geodesics.

(vi) The fact that there is a local minimum of the effective potential
for both null and timelike geodesics ensures that there are stable circular
orbits with radius between $r_{2}$ and $r_{1}$.

\vskip 0.3cm

(b) $ Q^6= \frac{27}{64}M^4$  

Here, equations (\ref{eq:711})--(\ref{eq:713}) reduce to
$r_+=r_- = \frac{4Q^2}{3M\alpha}.$
We then see that for the extremal static charged black string the general
features of the geodesics are the same as for the non-extremal case discussed
above. All the properties listed in paragaphs (i)--(vi) still hold. There
are, however, some peculiarities.

All geodesics with $E^2=0$ must have $r=\frac{4Q^2}{3M\alpha}$. 
In fact, the right hand side of (\ref{eq:709}) is null for 
$r=\frac{4Q^2}{3M\alpha}$ and is negative for all other values of $r$. 
A geodesic like this can represent a helical line ($P^2\neq0$, $L^2\neq 
0$), a circle ($P^2=0$, $L^2\neq 0$), a strait line along the $z$ 
direction ($P^2\neq0$, $L^2=0$), or even a ``radial'' geodesic with 
$r=r_+$ ($P^2=0$, $L^2= 0$). It is worth to note that even though 
$E=0$, it follows from (\ref{eq:707}) that $\dot{t}$ may be different 
from zero for particles on the horizon, since $B=0$ there. Such 
geodesic orbits are stable against small radial perturbations.

\vskip 0.5cm

{\bf VII.2 The rotating case}

As we have mentioned before, we constrain ourselves to the rotating
black string case, where $a$ is restricted by $0 < a^2\alpha^2\leq
\frac{2}{3} - \frac{128}{81} \frac{Q^6}{(1-\frac{1}{2}
\alpha^2a^2)^3M^4}$. Due to the similarity among equation
(\ref{eq:704}) and (\ref{eq:708}) the behaviour of  radial
geodesic motion in both static and rotating black string
spacetimes are also very similar, as we shall see below.

\vskip 0.3cm
(a) $0 < a^2\alpha^2< \frac{2}{3} - \frac{128}{81} 
\frac{Q^6}{(1-\frac{1}{2}
\alpha^2a^2)^3M^4}$.

This condition represents the non-extremal black string where polynomial 
$B$ has two real 
positive roots, the two horizons $r_{+}$ and $r_{-}$, given respectively by
(\ref{eq:4112}) and (\ref{eq:4113}). It is negative in the region $r_{-}<
r<r_{+}$, otherwise it is positive.

Then, we see from (\ref{eq:704}) that as in the static case there are two
turning points $r_{2}$ and $r_{1}$, with $r_{2}>r_{+}$ and $r_{1}<r_{-}$,
and the allowed 
region for the radial coordinate of geodesic particles is between
$r_{1}$ and $r_{2}$.
Again,  for null geodesics with $E^2 -\alpha^2L^2 -P^2 \geq 0$ the turning
point $r_{2}$ is at infinity.
Here, however, the turning points coincide with the horizons for timelike
and null geodesics whose motions are such that $c_1=0$, i.e.,
the radial coordinate of the geodesics must be between the horizons
if the energy and angular momentum of the particle satisfy
$E\sqrt{1-\frac{\alpha^2a^2}{2}}-\alpha^2La=0$.

We then start studing the motion of null particles. Once more the important
equation to analyze is (\ref{eq:704}) which we rewrite here (with
$\varepsilon^2=0$)
\begin{eqnarray}
\alpha^4r^4\dot{r}^2&=&(E^2-P^2 -\alpha^2L^2)\alpha^4r^4+ 4M
\left(P^2(1-\frac{3\alpha^2a^2}{2})+\alpha^2c_{0}^2\right)\alr- \nonumber\\
& &\frac{4Q^2}{1-\frac{\alpha^2a^2}{2}}
\left(P^2(1-\frac{3\alpha^2a^2}{2})+\alpha^2c_{0}^2\right)\, , \hfill
                   \label{eq:718}
\end{eqnarray}
where $c_{0}^2$ is given by (\ref{eq:705}).
Then, comparing (\ref{eq:718}) with (\ref{eq:709}) and (\ref{eq:715}) we 
can take the following conclusions.

(i) All what was said in VII.1 (i) also holds for the rotating
black string. Null particles can escape to infinity only if they have more
than the escape energy, i.e., if $E^2 \geq\alpha^2 L^2+P^2$.

(ii) The conclusions in VII.1 (ii) are also valid here. Thus,  null geodesics
with $E^2 <\alpha^2 L^2+P^2$ are bounded between a maximum and a minimum
radial distance and
cross the horizons an infinite number of times. These kind of geodesics
do not reach the singularity, since in this case $c_0^2\neq 0$ (see (iii)
below).

(iii) Only null geodesics with $P^2=0$ and $c_{0}^2=0$ can reach the
singularity.
These conditions imply $E^2- \alpha^2L^2 > 0$. Such geodesics describe
the motion of particles which do not travel in the $z$ direction and
whose angular momentum and energy are related by $\frac{L}{E}= a/
\sqrt{1-\frac{\alpha^2a^2}{2}}=\frac{\omega}{\alpha^2\gamma}$, where
$\gamma$ and $\omega$ are given respectively by (\ref{eq:4105}) and
(\ref{eq:4106}). Such geodesic motion satisfies the relation
$\frac{d\varphi}{dt}= \frac{\alpha^2L}{E}= \frac{\omega}{\gamma}$, which
means that this kind of geodesics rotates with the same angular velocity
of the black string. They are radial geodesics (with respect to the
rotating string) in the sense that the angular coordinate
$\bar{\varphi}$ defined in equation (\ref{eq:4110}) is constant
($\frac{d\bar{\varphi}}{d\tau} =0$). In this case, the radial equation can
be integrated producing the simple relation (see also (\ref{eq:716}))
\begin{equation}
r  = r_0 \pm \frac{E}{\sqrt{1-\frac{\alpha^2a^2}{2}}}\tau = r_0 \pm
\frac{L}{a}\tau\, ,
\nonumber \\
\end{equation}
which describes either a geodesic coming from large values of $r$ and
spiralling to the singularity in a finite proper time 
or a spiral line starting at $r=0$ and extending to infinity.

(iv) There is also a particular case similar to VII.1 (iv). If
$E^2 =\alpha^2L^2 +P^2$ the equation for $\frac{dr}{d\tau}$ yields (see
also (\ref{eq:717}))
\begin{equation}
\pm 15A\sqrt{\alpha M}\,\tau+{\rm const.} =\left(3r^2+{4r_1}r
+8r_1^2\right)\sqrt{r-{r_1}}\, ,
\nonumber \\
\end{equation}
where $\alpha r_{1}= \frac{Q^2}{M\left(1-\frac{\alpha^2a^2}{2}\right)}$ and
$A= a\alpha^2L-\sqrt{(P^2+\alpha^2L^2)\left(1-\frac{\alpha^2a^2}{2}\right)}$.
Such equation describes either spiralling null geodesics
that start at $r_{1}$
and reach infinite radial distances, or geodesics that start at
$r=+\infty$ and spiral towards $r_{1}$ and then return to infinity.

(v) Regarding timelike geodesics we have to consider the full radial
equation (\ref{eq:704}). The general features are much the same as for null
particles with $E^2-\alpha^2 -P^2<0$ discussed in VII.1 (ii) above.
That is to say, the motion of any timelike geodesic
is bounded within the region $r_{1} \leq r\leq r_{2}$. Then, an
uncharged timelike particle emited by the black string reachs the maximum
distance at $r=r_{2}$. Then is pulled back crosses inwards the two
the horizons $r_{+}$ and $r_{-}$ and reaches a minimum distance 
at $r=r_{1}$. There, it is scattered by a potential barrier
to $r_2$, after crossing $r_+$ and $r_-$ of a new universe in the Penrose 
diagram (see figure 4.2). 
Thus, the geodesic has an infinite length,  it crosses an 
infinite number of times the two horizons. 

(vi) From the form of the effective potential it also follows that 
there are stable circular geodesics.

\vskip 0.3cm

(b) $a^2\alpha^2=\frac{2}{3}
-\frac{128}{81} \frac{Q^6}{(1-\frac{1}{2}\alpha^2a^2)^3M^4}$.

This corresponds to the the extremal rotating black string.
The particularities of geodesic motion here
are  similar to the extremal static case studied in VII.1(b). As we have
seen, all particles whith $c_1^2=0$, have their motion restricted to the
region bounded by the two horizons. Then, since the two horizons 
coincide here,
the only possible geodesics with $E\sqrt{1-\frac{\alpha^2a^2}{2}}=
\alpha^2La$ are the ones that have constant radial coordinate
$r= \frac{4Q^2}{3M\alpha(1-\frac{\alpha^2a^2}{2})}$.
For $c_1^2\neq 0$ the behaviour of the radial coordinate of the geodesic
motion is the same as in the non-extremal case.

Finally, let us recall that there are other three special cases considered
in section IV.1.
Namely, (1) the naked singularity case (cf. IV.1(iii)); (2) the topological
singularity case (cf. IV.1(iv));
and (3) the other black string case (cf. IV.1(v)).
In all theses cases the asymptotic behaviour of geodesics for large values
of $r$ is
the same as for the rotating black string spacetime. The most important
particularity is that in (2) and (3) some particular timelike 
(in addition to the null) geodesics can also reach the singularity.

\vspace*{1.cm}
\addtocounter{chapter}{1}
\setcounter{equation}{0}
\setcounter{figure}{0}
\vspace{10pt}

{\bf  VIII. Dimensional reduction and the 3D black hole}

In this section we discuss the connection between the 4D black string
and a 3D charged and rotating black hole through the dimensional
reduction of actions (\ref{eq:21})  under suitable
conditions as seen bellow. Such a procedure was used in ref.
\cite{lemos1} in order to define the parameters of the 3D black hole and
it was then possible to postulate  mass and angular momentum per unit length 
for the 4D
black string by simply taking the values of the mass and angular momentum 
from the 3D spacetime.
Repeating the dimensional reduction formalism, but now for the charged
case, we are now able to show explicitly that the mass, angular
momentum and electric charge of the 3D black hole are exactly the line
density quantities of the 4D black string defined in section III.

Consider a  4D spacetime metric admitting one spacelike Killing vector,
$\dspst{\partial\over{\partial z}}$. In  such a case the 4D metric may be
written in the form
\begin{equation}
ds^2= (g_{mn}-X_mX_n) dx^{m}dx^{n} + 
2g_{mn} e^{-4\phi}X^{m}dx^{n}dz + e^{-4\phi}dz^2\, ,
     \label{eq:81}
\end{equation}
where the 3D metric $g_{mn}$ ($m,n=0,1,2$) and $\phi$ and $X^{m}$ are 
metric functions independent of $z$.
The spacetime we are considering here (see e.g.  equations 
(\ref{eq:21}) and (\ref{eq:32})) is a
particular case of (\ref{eq:81}) where $X^{m}=0$. 
After the splitting (\ref{eq:81}) and taking  $X^{m}=0$, 
the electromagnetic action can be put 
in the following  suitable form 
\begin{equation}
S_{\rm em} = \frac{1}{8\pi}\int     d^{4}x\sqrt{-g}
\left( {H^{mn}H_{mn}\over2}+ e^{4\phi}{B^{m}B_{n}}\right) ,
                                \label{eq:82}
\end{equation}
where we have introduced the definitions
\begin{equation}
 H_{mn} \equiv F_{mn}, \, \quad\quad  B_{m}\equiv F_{3m} = -F_{m3},\,
     \label{eq:83a}
\end{equation}
\begin{equation}
 F_{\mu\nu} \equiv \left(\begin{array}{cc}
         H_{mn} & -B_{n}\\
      B_{m}  &0 \end{array}\right)  , 
                                 \label{eq:83b} 
\end{equation}
with $H^{mn}\equiv g^{mp}g^{nq}H_{pq}$, etc.
In the above equations, greek indices run from $0$ to $3$ and 
roman indices from $0$ to $2$. 

For our purposes in
this paper we may also put $B^{m}=0$ since it vanishes identically for the
charged stationary black string. For instance, according to 
(\ref{eq:26}), the electromagnetic potential assumes the form $ A_{\mu} =
- \gamma h(r) \delta_{\mu}^{0} +\frac{\omega}{\alpha^2}h(r)
\delta_{\mu}^{2}$, where $h(r) = \frac{2\lambda}{\alr}$, from which
follows that $B_{m} \equiv F_{3m} = 0$.

Then after dimensional reduction we obtain the following 3D actions
\begin{eqnarray}
S_{\rm em}& =& -\frac{1}{16\pi}\int{\!\! d^{3}x\sqrt{-g}e^{-2\phi}
\left(H^{mn}H_{mn} \right)} \, , \label{eq:84}\\
S&=& \frac{1}{16\pi G}\int{\!\! d^{3}x\sqrt{-g}e^{-2\phi}(R-2\Lambda)}\, .
\label{eq:85}
\end{eqnarray}
By varying the action $S + S_{\rm em}$ with respect to $g_{mn}$, $\phi$ and
the electromagnetic potential $A_{m}$ one obtains equations of motions
identical to 4D Einstein-Maxwell equations with one Killing vector 
$\frac{\partial}{\partial z}$. Imposing stationarity 
and axial symmetry in the 3D theory one finds a 3D
charged black hole solution which generalizes the stationary 3D black hole
of \cite{lemos1} and whose metric can be written in the form:
\begin{eqnarray}
&ds^{2} = -\left(\asr -\frac{4M(1-\frac{a^2\alpha^2}{2})}{\alr} + 
\frac{4Q^2}{\asr}\right) dt^2 + &
\nonumber \\
&-\frac{4aM\sqrt{1-\frac{a^2\alpha^2}{2}}}{\alr}\left(1-
\frac{Q^2}{M(1-\frac{a^2\alpha^2}{2})\alr}\right) 2dt d\varphi +&
\nonumber \\
&+ \left(\asr -\frac{4M(1-\frac{3}{2}a^2\alpha^2)}{\alr} + 
\frac{4Q^2}{\asr} 
\frac{(1-\frac{3}{2}a^2\alpha^2)}{(1-\frac{a^2\alpha^2}{2})}
\right)^{-1} dr^2 +&
\nonumber \\
&+ \left[r^2 + \frac{4Ma^2}{\alr}\left(1-
\frac{Q^2}{(1-\frac{a^2\alpha^2}{2})M\alr}\right)\right] d\varphi^2\, , &
\nonumber\\
& e^{-2\phi}= d_0\alr,&
                             \label{eq:85b}
\end{eqnarray}
where $d_0$ is a constant which here we take equal to unity, $d_0=1$. 
To find the mass and angular momentum in 3D, and how it relates to the 
mass and angular momentum per unit length in 4D, we write equation 
(\ref{eq:85b}) in the canonical form , 
\begin{eqnarray}
&ds^2 =-{N^{0}}^2 dt^{2}+R^2(N^{\varphi}dt+d\varphi)^{2}+
\frac{dR^{2}}{f^2}\, ,&
\nonumber\\ & e^{-2\phi} = \alr\ , \label{eq:86}
\end{eqnarray}
where  $N^{0}$, $N^{\varphi}$, $R$, $f^2$ are defined in equation 
(\ref{eq:32}), or else can be taken directly from (\ref{eq:85b}). 
From (\ref{eq:84}), (\ref{eq:85}) and (\ref{eq:82}) we get an action 
written in the Hamiltonian formalism, 
\begin{eqnarray}
&S =-{1\over {8G}}\int\! dt\, \left\lbrace N\left\lbrack {1\over2}e^{-2\phi}
 R^{3}{\left({N^\varphi}_{,R}\right)^{2}\over N^2}+e^{-2\phi}\left(f^{2}
\right)_{,R}
\left(1-2R{d\phi\over dR}\right)- \right.\right.& \nonumber\\
&\left.\left. -4f^2\left( e^{-2\phi} R {d\phi\over dR}\right)_{,R} +
2e^{-2\phi}R\Lambda\right\rbrack +N^\varphi\left(e^{-2\phi}R^3
{{N^\varphi}_{,R}\over N} \right)_{,R} \right\rbrace dR &  \nonumber \\
&+ S_{\rm em} + B\, ;&\hfill \label{eq:87} \\
&S_{\rm em}  \equiv  \int\! dt\, \left\lbrace N\left\lbrack
 {e^{-2\phi} \over 4R}\left(e^{4\phi}{\E}^{2}+\left({A_{2}}_{,R}
\right)^{2}\right) \right\rbrack
 +N^\varphi\left(\frac{{\E}{A_{2}}_{,R}}{2}\right)
+ \frac{A_{0}{\E}_{,R}}{2}\right\rbrace dR &\nonumber
\end{eqnarray}
where $N\equiv{N^0\over f}$, $N^\varphi$ and $A_{0}$  are Lagrange
multipliers, and  $B$ is a surface term.

The next step is analyzing each surface term that follows by varying the
action (\ref{eq:87}) with respect to $\phi$, $f^2$ and $A_{2}$  and
their conjugate momenta \cite{rt,btz,bhtz}. It follows that the surface
terms that comes from the gravitational action for the charged 3D black hole
do not change with respect to the uncharged 3D black hole. Moreover, 
apart from the term $A_{0}{\E\over2}$, 
the new surface terms in the total action coming from the 
electromagnetic field also vanish when $R\longrightarrow\infty$ . 
That is to say, the 
conjugate quantities to $N$ and $N^{\varphi}$ which correspond respectively
to (ADM) mass and angular momentum of the charged 3D black hole are the same
as if the black hole was not charged and are given respectively by
the right hand side of (\ref{eq:34}) and (\ref{eq:35}).

The only new surface term which survive at infinity is $A_{0}
\delta({\E\over2})$ from which we find the electric charge of the black hole
(as the conjugate quantity to the Lagrange multiplier $A_{0}$):
\begin{equation}
 Q\equiv{1\over2}(\delta{\E})_{R=\infty}= \gamma\lambda,
                             \label{eq:89}
\end{equation}
which is exactly the same as (\ref{eq:39}).

From these results we see that the mass, angular momentum and charge of the 3D
black hole are exactly the mass $M$, angular momentum $J$ and charge $Q$
per unit length of the (4D) black string. However, 
this is valid only for $d_0>0$ in (\ref{eq:85b}).
Notice also that $M$ and $J$
given in (\ref{eq:34}) and (\ref{eq:35}) are identical to the mass and angular
momentum of 
\cite{lemos1}, apart from the normalization factor $G= {1\over8}$.

Also note that whenever $\phi=0$ in (\ref{eq:85}) one obtains the 
Einstein's 3D gravity and the BTZ 3D black hole follows. This black 
hole can therefore be related to a 4D cylindrical black hole in a 
simple manner, although the corresponding 4D black string has a non-zero 
energy-momentum tensor \cite{lemos3}.

\vspace*{1.cm}
\addtocounter{chapter}{1}
\setcounter{equation}{0}
\setcounter{figure}{0}
\vspace{10pt}

{\bf  IX. Conclusions}

We have found a solution for a charged rotating black string in General
Relativity. The maximal analytical extension of the charged solutions,
studied in sections IV, have shown some similarities with the
Kerr-Newman family of black holes.  There is an ergosphere, there are
event and Cauchy horizons, closed timelike curves inside the Cauchy
horizon and timelike singularities. There are regions, such as the
region behind the past event horizon (also appearing in the Kerr-Newman
solutions),  which will be covered when one performs complete
gravitational collapse of some cylindrical matter in an appropriate
background.
 
The maximal analytical extensions of the uncharged solutions do not
resemble so much the Kerr solution. The rotating uncharged case shows
similarities with the Schwarzschild black hole. For instance, the
singularity is spacelike.
 
It was also shown that this black string solution corresponds to a 
3D black hole. Thus, one has a framework in which it is possible 
to relate lower dimensional results with 4D General Relativity
(see also \cite{lemos1,lemos2,sakleber,lemos4,lemossa}).

\vspace*{1.cm}
\noindent {\large\bf Aknowledgements - } 
VTZ would like to thank the  Departamento
de Astrof\'{\i}sica of the  Observat\'orio Nacional - CNPq
(Rio de Janeiro) for hospitality and financial support 
while part of this work was being done.

\vfill\eject 
\centerline{Figure Captions}

Figure 4.1 -- The five regions and lines which 
yield solutions of different nature are shown. 

Figure 4.2 -- The Penrose diagram representing the non-extreme charged 
rotating cylindrical black hole. The double line at $r=0$ indicates 
a scalar polynomial singularity. 

Figure 4.3 -- The Penrose diagram for the extremal charged rotating 
cylindrical black hole.

Figure 4.4 -- The Penrose diagram for the charged rotating 
naked singularity. 

Figure 4.5 -- The Penrose diagram for the charged rotating 
null singularity. 

Figure 4.6 -- The Penrose diagram for the charged rotating 
solution with one horizon. 

Figure 5.1 -- The Penrose diagram for 
the rotating uncharged black hole. 

Figure 5.2 -- The Penrose diagram for the extremal 
rotating uncharged black hole. The singularity at $r=0$ is
non-polynomial. 

Figure 5.3 -- The Penrose diagram for the rotating 
uncharged naked singularity.

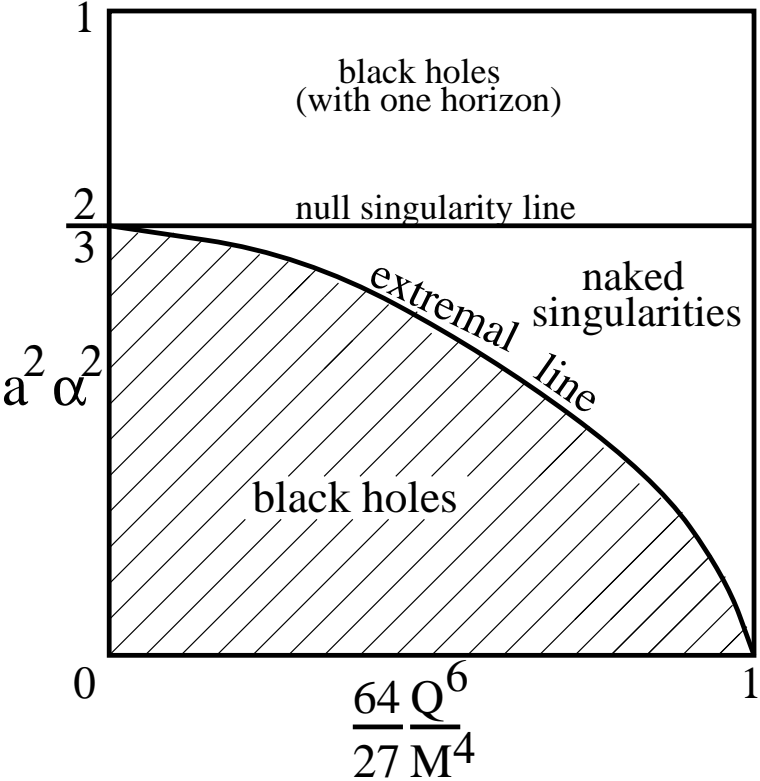
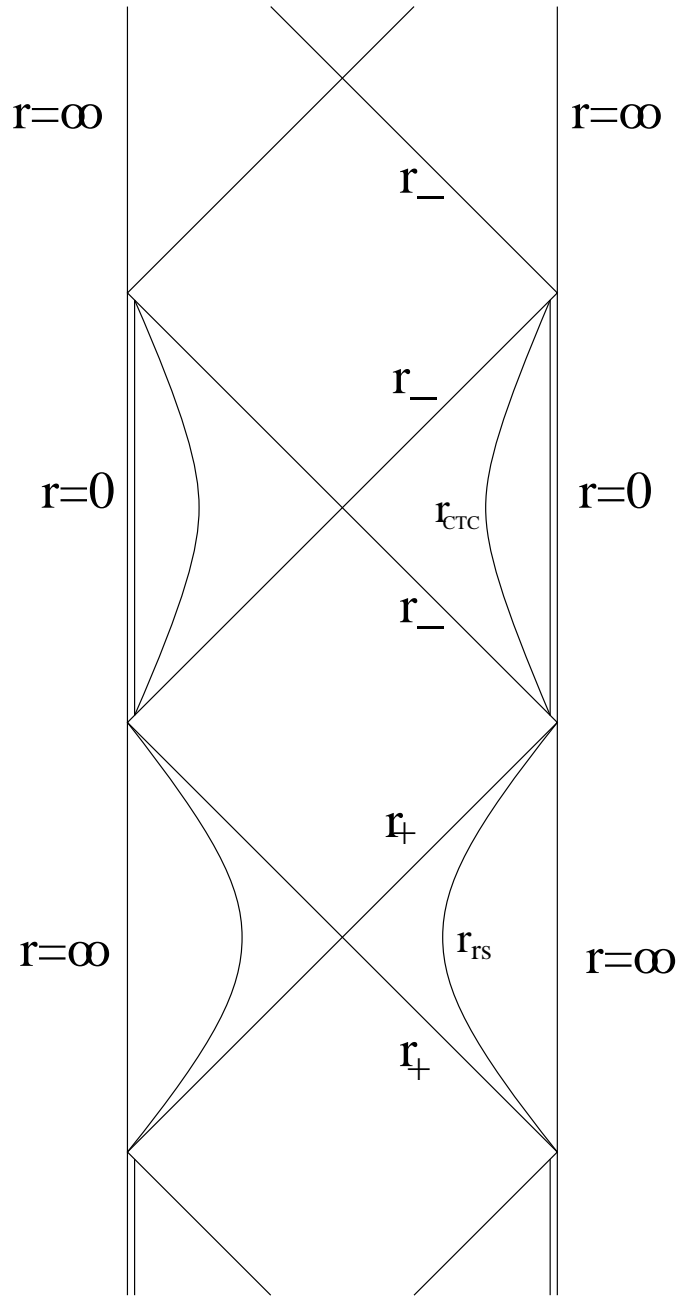
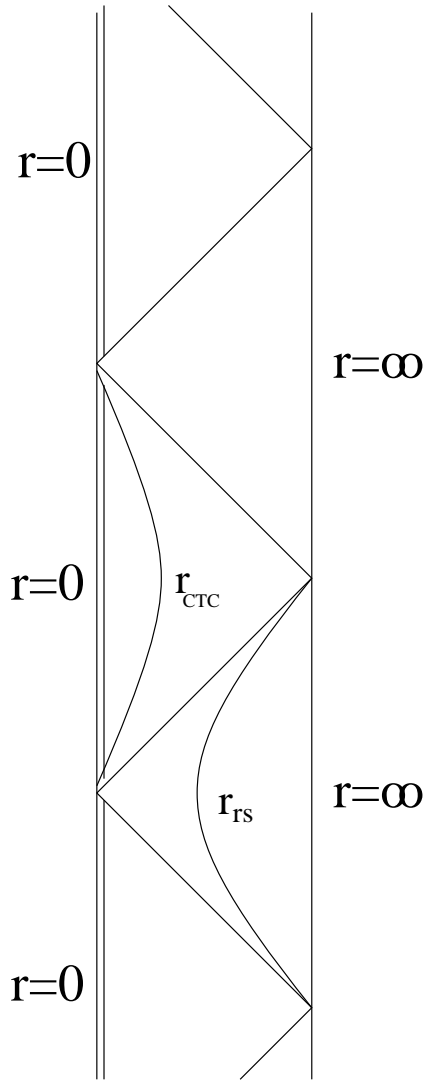
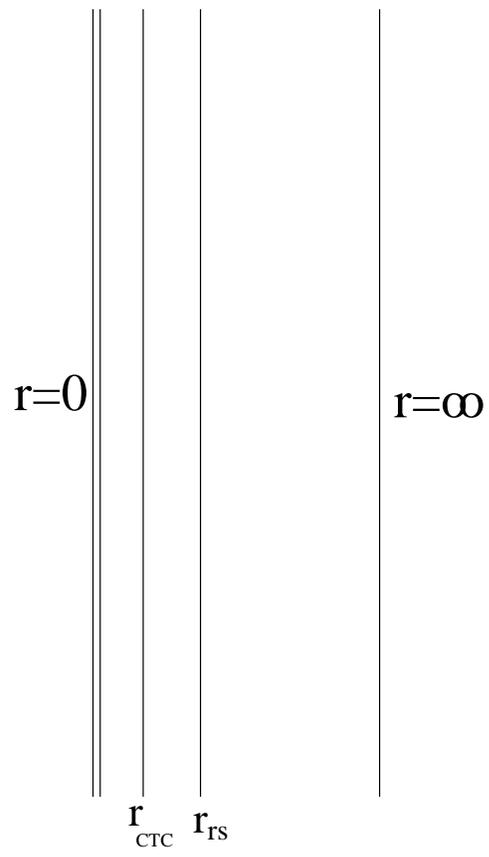
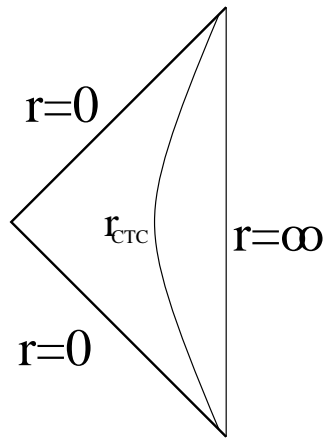
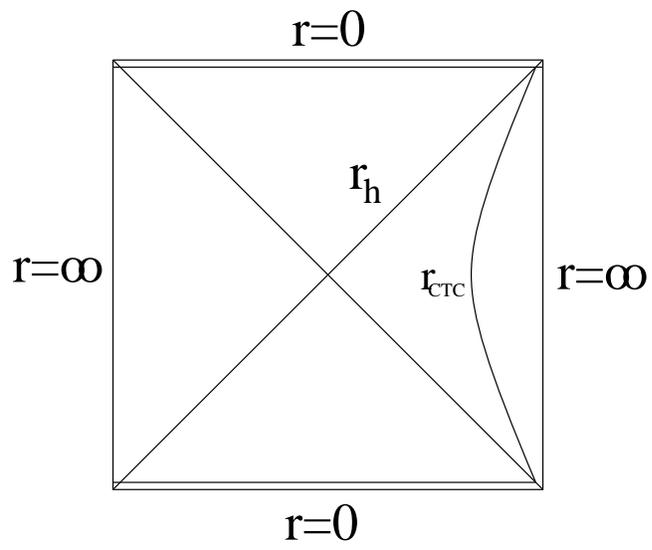
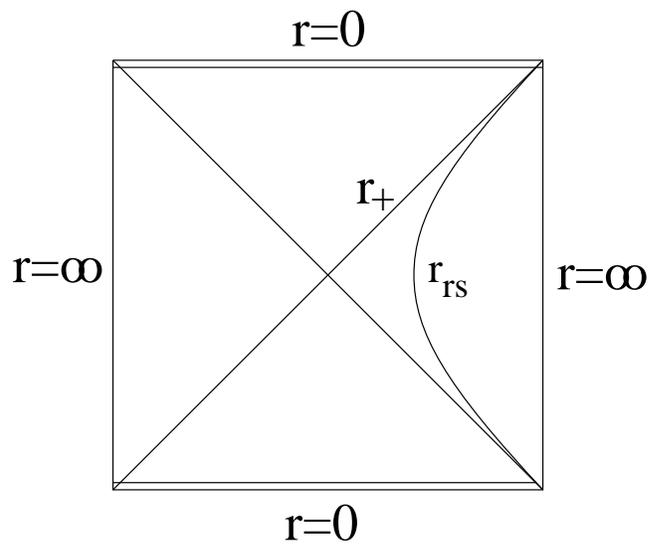
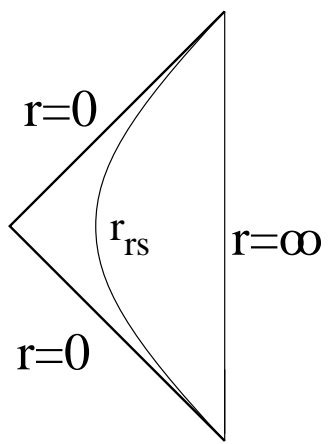
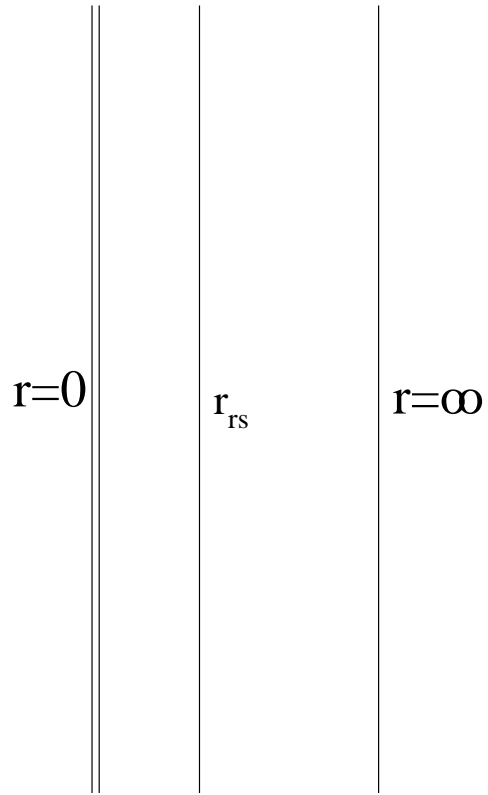
\end{document}